\documentclass[aps,prb,twocolumn,showpacs, amsmath,floatfix]{revtex4}
\usepackage{epsfig}
\begin{document}  
%@twocolumnfalse\endcsname

\title{Theory of Structural Response to Macroscopic Electric Fields in
Ferroelectric Systems}
\author{Na Sai, Karin M. Rabe and David Vanderbilt}
\affiliation{Department of Physics and Astronomy, 
Rutgers University, Piscataway, NJ 08854-8019}
% \date{\today}
\date{May 20, 2002}

\begin{abstract}
We have developed and implemented a formalism for computing the structural response
of a periodic insulating system to a homogeneous static electric
field within density-functional perturbation theory (DFPT). 
We consider the thermodynamic potentials $E({\bf R},\eta,{\cal E})$ and
$F({\bf R},\eta,{\bf P})$, whose minimization with respect to the
internal structural parameters ${\bf R}$ and unit cell strain $\eta$
yields the equilibrium structure at fixed electric field $\cal E$ and
polarization ${\bf P}$, respectively.  First-order expansion of $E({\bf
R},\eta,{\cal E})$ in ${\cal E}$ leads to a useful approximation in which
${\bf R}({\bf P})$ and $\eta({\bf P})$ can be obtained by simply minimizing the zero-field internal
energy with respect to structural coordinates subject to the constraint of
a fixed spontaneous polarization ${\bf P}$.  To facilitate this
minimization, we formulate a modified DFPT scheme such that the computed
derivatives of the polarization are consistent 
with the discretized form of the Berry-phase
expression. We then describe the application of this
approach to several problems associated with bulk and short-period
superlattice structures of ferroelectric materials such as BaTiO$_3$ and
PbTiO$_3$.  These include the effects of compositionally broken inversion
symmetry, the equilibrium structure for high values of polarization,
field-induced structural phase transitions, and the lattice
contributions to the linear and the non-linear dielectric constants.
\end{abstract}

\maketitle
%\vskip2pc] 

%===========================================================================
\section{Introduction}
\label{sec:intro}
%==========================================================================
As the usefulness of density-functional theory (DFT) for the study
of dielectric materials is now well established, one might imagine
that calculations of crystalline insulators in the presence of a
homogeneous macroscopic electric field should be routine.  On the
contrary, the presence of an electric field introduces several
severe difficulties.\cite{NV,NG}  The electric potential acquires a
term that is linear in the spatial coordinates, thus violating the
periodicity condition underlying Bloch's theorem and acting as a
singular perturbation on the electronic eigenstates.  Moreover, in
principle there is no longer a well-defined ground state for the
electrons in a solid in a macroscopic electric field because the
energy of the system can be always lowered by transferring
electrons from the valence band in one spatial region to the
conduction band in a distant region.

One way around these difficulties is to make use of density-functional
perturbation theory (DFPT),\cite{BGT,GonzeVig,Gonze95} which
provides a framework for calculating the perturbative response to
infinitesimal electric fields (as well as to atomic displacements
and strains).  DFPT has been widely adopted for many studies of the
dielectric and piezoelectric properties and dynamic effective
charges of dielectric materials.  However, being a perturbative
approach, the method is not capable of treating a finite electric
field directly.

A more direct attack on the finite-field problem was made by Nunes
and Vanderbilt,\cite{NV} who showed that a real-space
Wannier-function representation could be used to represent the
electronic system in the presence of a finite electric
field.\cite{NV} In this approach, one minimizes a total-energy
functional of a set of field-dependent Wannier functions for a
periodic system at fixed electric field.  Alternatively, the
minimization can also be performed at fixed polarization via a
Legendre transformation of the energy functional with the electric
field treated as a Lagrange multiplier.  The approach was
implemented in the DFT context by Fernandez {\it et al.},
\cite{Fernandez} but proved too cumbersome to find widespread
utility.

In the present paper we propose a new scheme for the treatment of
a dielectric system in a static homogeneous electric field.  Our
scheme is based on a low-order truncation of the DFPT perturbative
expansion in electric field, and the use of this truncated
expansion to extrapolate to finite electric field.  A key feature
of our approach is that, while we keep only low orders in the
expansion in electric field, we effectively keep all orders of
expansion in the structural degrees of freedom.  We demonstrate
that even a first-order truncation of the electric-field
perturbation provides a very useful and practical scheme.  In this
context the electric field simply couples to the
zero-field polarization, so that the latter plays a central role
in our formulation.  In fact, it is rather natural to formulate our
approach in terms of a constrained minimization procedure in which
the DFT energy functional is minimized over all structural degrees
of freedom subject to a constraint on the value of the
polarization.  This allows a two-step approach in which one
first maps out the energy surface as a function of polarization in
the DFT framework, and then uses this energy surface, augmented by
the coupling to the electric field, to obtain the ground-state
structure in the presence of the field.  We will show that
essentially no additional approximations are needed beyond the
first-order truncation of the free-energy expansion in electric
field.  We will also show that the methodology can be extended to
second order (or, in fact, to any desired order) in the electric
field.

Before proceeding, we should acknowledge an additional theoretical
subtlety associated with the correct choice of exchange-correlation
functional in the electric-field problem.  Gonze, Ghosez and Godby
\cite{GGG95} (GGG) argued that the exact Kohn-Sham exchange-correlation
energy functional should have a dependence on the macroscopic
polarization and formulated a ``density-polarization functional
theory'' in which there is generally an exchange-correlation
contribution to the Kohn-Sham electric field.  \cite{GGG95}  While
this was an important formal development that subsequently
received much attention,\cite{Resta96,Martin97,GGG97} it has not
yet led to an improved practical exchange-correlation functional.  We
thus restrict ourselves here to the usual LDA exchange-correlation
functional where, because of the locality of the central approximation,
the subtleties identified by GGG do not arise.
 
This paper is organized as follows. In the next section, we present
our formalism for computing the structural response of an
insulating system to an electric field.  Some details of
the implementation are presented in Sec.~\ref{sec:Method},
including details of our minimization procedure, a discussion of
modifications that we made to the DFPT procedure to achieve
compatibility with the discretized Berry-phase polarization
formula, and a description of the technical details of the  {\it
ab-initio} pseudopotential calculations.  Then, in
Sec.~\ref{sec:Appli}, we present several sample applications of our
method.  In Sec.~\ref{sec:ISB}, we show that it provides an
alternative approach to the study of short-period ferroelectric
superlattice structures with broken inversion symmetry.\cite{Sai}
In Sec.~\ref{sec:BT} we present a study of the dependence of the
internal structural parameters
of BaTiO$_3$ on polarization. In 
Sec.~\ref{sec:Dielectric} we show that our method provides a
straightforward way of computing the dielectric susceptibilities
and piezoelectric coefficients as functions of the electric
field, thus allowing an estimation of the non-linear dielectric and
piezoelectric response in a ferroelectric system. Finally, in
Sec.~\ref{sec:SPT}, we consider a case in which a full
three-dimensional treatment of the polarization and the structural
distortions is needed. Specifically, we model the polarization rotation and
structural phase transitions induced by the application of a macroscopic
electric field to a model ferroelectric system and relate our
results to recent experiments in PZN-PT.\cite{Cox} Finally in
Sec.~\ref{sec:Summary}, we summarize our work and discuss the prospects
for future applications of our approach.

%=======================================================================
\section{Formalism}
\label{sec:forma}
%=======================================================================

Our goal is to investigate the effect of a homogeneous
static electric field on the structure and polarization of polar
insulators, including systems with a nonzero spontaneous
polarization.  In addition to an efficient approach for
computation, we also aim towards a formulation that readily allows
an intuitive understanding of the effects of the field. As will
become clear below, this will lead us to a formulation in which the
polarization plays an especially prominent role.

%------------------------------------------------------------------------
\subsection{Case of a single minimum}
\label{sec:SingleMin}
%------------------------------------------------------------------------

Let ${\bf {\cal E}}$ be the macroscopic electric field, ${\bf R}$ the
atomic coordinates, and $\eta$ the lattice strain, and assume that
the total energy per unit cell $E({\bf R},\eta,{\bf {\cal E}})$
has a single local minimum of interest in the $({\bf R},\eta)$ space for
given ${\bf {\cal E}}$.  (This restriction is normally appropriate for a
paraelectric material, but not for a ferroelectric one.  The existence
of multiple local minima in the latter case calls for a more careful
discussion, which is deferred to the following subsection.)  We let
\begin{equation}
E({\bf {\cal E}})=\min_{{\bf R},\eta} {E({\bf R},\eta,{\bf {\cal E}})}
\label{eq:minE}
\end{equation}
and denote the location of the minimum by
${\bf R}_{\rm eq}(\cal E)$ and $\eta_{\rm eq}(\cal E)$.
The polarization
${\bf P}({\bf R},\eta,\cal E )$, the thermodynamic conjugate of
$\cal E$, can then be obtained from the expression
${\bf P}({\bf R},\eta,{\cal E}) = -[d E({\bf R},\eta, {\cal E})/ d{\cal
E}]_{{\bf R},\eta}$ and ${\bf P}(\cal E )$ obtained by
evaluating
${\bf P}({{\bf R}_{\rm eq}({\cal E}),\eta_{\rm eq}(\cal E),\cal E})$,
or equivalently, $-{d E({\cal E})/d{\cal E}}$.

We can recast this minimization into a form in which the polarization
is more central. Viewing $E({\bf R},\eta,{\bf {\cal E}})$ as a
thermodynamic potential that minimizes to equilibrium values of ${\bf
R}$ and $\eta$ at fixed ${\cal E}$ leads naturally via a Legendre
transformation to a thermodynamic potential $F({\bf R},\eta,{\bf P})$
that minimizes to equilibrium values of ${\bf R}$ and $\eta$ at fixed
${\bf P}$:
\begin{eqnarray}
F({\bf R},\eta,{\bf P}) &=& \min_{\lambda}[E({\bf R},\eta,\lambda)+
\lambda \cdot\bf P]\nonumber\\
&=& E({\bf R},\eta,\lambda({{\bf R},\eta, {\bf P}}))+
\lambda({{\bf R},\eta, {\bf P}}) \cdot {\bf P}
\label{eq:Fdef}
\end{eqnarray}
with $\lambda({{\bf R},\eta, {\bf P}})$ being the value at the
minimum. This is equivalent to ${\bf P}({\bf R},\eta,{\lambda})={\bf
P}$, that is, $\lambda({{\bf R},\eta, {\bf P}})$ is the value of the
macroscopic field necessary to produce polarization ${\bf P}$ at given {\bf
R} and $\eta$.

We then define the function
\begin{eqnarray}
F({\bf P}) &=& \min_{{\bf R},\eta}{F({\bf R},\eta,{\bf P})}\nonumber\\
&=&F({\bf R}_{\rm eq}({\bf P}),\eta_{\rm eq}({\bf P}),{\bf P})
\label{eq:FP}
\end{eqnarray}
with ${\bf R}_{\rm eq}({\bf P})$ and $\eta_{\rm eq}({\bf P})$ being the
values at the minimum.  These structural parameters
${\bf R}_{\rm eq}({\bf P})$ and $\eta_{\rm eq}({\bf P})$ are in fact equal
to ${\bf R}_{\rm eq}({\cal E})$ and $\eta_{\rm eq}({\cal E})$, the
structural parameters defined by minimizing
$E({\bf R},\eta,{\cal E})$ at the corresponding fixed
${\cal E}= \lambda({\bf R}_{\rm eq}({\bf P}),\eta_{\rm eq}(\bf P), {\bf P})$.
The polarization at this extremum is, as expected,
\begin{eqnarray}
{\bf P}({\bf R}({\cal E}),\eta({\cal E}),{\cal E})&=&\nonumber\\
{\bf P}({\bf R}_{\rm eq}({\bf P}),\eta_{\rm eq}({\bf P}),
\lambda({\bf R}_{\rm eq}({\bf P}),\eta_{\rm eq}(\bf P), {\bf P}))&=&{\bf
P} \, .
\end{eqnarray}

Finally, we re-express the original minimization as
\begin{eqnarray}
E({\cal E})&=&\min_{\bf P}[F({\bf P})-{\cal E} \cdot {\bf P}] \, .
\label{eq:newEtot}
\end{eqnarray}
In this expression, the electric field $\cal E$ appears only in the
term $-{\cal E} \cdot {\bf P}$, and thus the effects of the field
can be completely understood by investigating the $\cal E$-independent
free energy landscape $F({\bf P})$.

%------------------------------------------------------------------------
\subsection{Case of multiple stationary points}
\label{sec:MultiMin}
%------------------------------------------------------------------------
In many cases of interest, the function $E({\bf R},\eta,{\bf {\cal E}})$
has several local minima, and the essential physics of the
problem is to map out the competition between these minima.  For
example, in a tetragonal ferroelectric like PbTiO$_3$, there are
six degenerate minima of this function at ${\cal E}=0$, and the
application of a nonzero $\cal E$ breaks the symmetry and establishes
one dominant domain orientation of the polarization.  However, it
may also be of interest to follow the behavior of the other local
minima, corresponding to metastable states, as well as
other stationary points of this energy surface.
For example, saddle points and local maxima of
$E({\bf R},\eta,{\bf {\cal E}})$ can correspond to stable states
for fixed {\bf P}.

In such cases, it is straightforward to generalize the previous
discussion by associating a label $(n)$ with each stationary
point of interest.  Thus the location of the stationary point is
denoted by ${\bf R}^{(n)}({\cal E})$ and $\eta^{(n)}({\cal E})$, and
$E^{(n)}({\cal E})=E({\bf R}^{(n)}({\cal E}),\eta^{(n)}({\cal E}),{\cal E})$
is the energy at the stationary point. 
The arguments of the
previous subsection carry over much as before.  The discussion
following Eq.~(\ref{eq:FP}) is modified by noting that the
minimization of $F({\bf R},\eta,{\bf P})$ with respect to ${\bf R}$
and $\eta$ at fixed $\bf P$ in Eq.~(\ref{eq:FP})
will always be associated with one of the stationary points of
$E({\bf R},\eta,{\bf {\cal E}})$ with respect to $\bf R$
and $\eta$ at the corresponding fixed $\cal E$; that is, 
${\bf R}_{\rm eq}({\bf P})={\bf R}^{(n)}({\cal E})$ and
$\eta_{\rm eq}({\bf P})=\eta^{(n)}({\cal E})$ for some $n$.

Finally, defining the global minimum $E({\cal E})=\min_n E^{(n)}({\cal E})$,
it is easy to see that Eq.~(\ref{eq:newEtot}) holds as before.

%------------------------------------------------------------------------
\subsection{Truncation of the expansion}
\label{sec:Truncation}
%------------------------------------------------------------------------

The central quantities appearing in the preceding subsections are
the energy $E({\bf R},\eta,{\cal E})$ and the polarization
${\bf P}({ \bf R},\eta,{\cal E})$ in a given electric field
$\cal E$.
Unfortunately, there is as yet no rigorous formulation of DFT for
the case of finite non-zero $\cal E$.  However, electric-field derivatives
of arbitrary order may be computed by the methods of
density-functional perturbation theory.  We thus expand in $\cal E$
around $\cal E$ = 0:
\begin{eqnarray}
E({\bf R},\eta,{\bf {\cal E}}) 
&=& E({\bf R},\eta, {\bf {\cal E}}=0)+
\sum_{\alpha}{\bf {\cal E}}_\alpha\left.\frac{\partial E({\bf
R},\eta,{\bf {\cal E}})}{\partial{\bf {\cal E}}_\alpha}\right|_{{\bf {\cal
E}}=0} \nonumber\\
&+&\frac{1}{2}\sum_{\alpha\beta}{\bf {\cal E}}_\alpha {\bf {\cal E}}_\beta 
\left.\frac{\partial^2E({{\bf R},\eta,{\bf {\cal
E}}})}{\partial{\bf {\cal E}}_\alpha\partial{\bf {\cal
E}}_\beta}\right|_{{\bf {\cal E}}=0}+\cdots
\label{eq:Eexp}
\end{eqnarray}

Carried to all orders in $\cal E$, this expansion is exact.  However,
for sufficiently small fields we can make the approximation of
truncating this sum
to define $E_i({\bf R},\eta,{\bf {\bf {\cal E}}})$ as the sum of the
first $i$+1 terms in Eq.~(\ref{eq:Eexp}), and ${\bf P}_i({\bf R},\eta
,{\bf {\bf {\cal E}}})$ as $\left.-({d E_i({\bf R},\eta, {\cal E})/
d{\cal E}})\right\vert_{{\bf R},\eta}$. Note that this
truncation is only in powers of $\cal E$, and that the dependence
on ${\bf R}$ and $\eta$ is preserved to all orders.

For many systems, it is already of interest to consider the
simplest case $i=1$, where
\begin{equation}
E_1({\bf R},\eta,{\cal E})=E({\bf R},\eta,0)-{\cal E}\cdot{\bf P}({\bf
R},\eta,0) \, ,
\label{eq:E1}
\end{equation}
\begin{equation}
{\bf P}_1({\bf R},\eta,{\cal E})={\bf P}({\bf R},\eta,0) \, .
\label{eq:P1}
\end{equation}
At this order, the resulting expression 
\begin{equation}
F({\bf P})=\min_{{\bf R},\eta,\lambda}[E({\bf
R},\eta,0)+{\lambda}\cdot({\bf P}-{\bf P}({\bf R},\eta,0))]
\label{eq:F1}
\end{equation}
can also be interpreted as one in which $\lambda$ appears
simply as a Lagrange multiplier implementing the constraint ${\bf P
}({\bf R},\eta,0) = {\bf P}$ in the set of equations 
that minimize $E({\bf
R},\eta,0)$ over ${\bf R}$ and $\eta$, i.e.,
\begin{eqnarray}
&\phantom{=}&\frac{\partial E({\bf R},\eta,0)}{\partial
R_{i\alpha}} - \sum_{\beta}\frac{\partial P_\beta({\bf R},\eta,0)}{\partial
R_{i\alpha}}\lambda_\beta = 0\nonumber\\
&\phantom{=}&\frac{\partial E({\bf R},\eta,0)}{\partial
\eta_{\mu}} - \sum_{\beta}\frac{\partial P_\beta({\bf R},\eta,0)}{\partial
\eta_{\mu}}\lambda_\beta = 0\nonumber\\
&\phantom{=}&P({\bf R},\eta,0) = {\bf P} \, .
\label{eq:mincond}
\end{eqnarray}
Details of the calculation are
described in the next section, and  all of the results 
reported in the following sections are obtained using the $i=1$
expressions.

Generalization of the formalism to order $i=2$  is
provided in Appendix~\ref{app:higherorder}.

%------------------------------------------------------------------------
\subsection{Relation to method of Fu and Cohen}
\label{sec:FuCohen}
%------------------------------------------------------------------------
In the remaining part of this section we discuss an earlier approach
proposed by Fu and Cohen (FC).\cite{FC} These authors carried 
out a first-principles investigation of the mechanism of rotation of the
polarization in BaTiO$_3$ by an applied electric field. Their approach
is similar to our $i=1$ case, but involves an additional approximation which we will describe by expressing their procedure in the notation established in this section.

The first step in their approach is the same as our $i=1$ case: to 
approximate $E({\bf R},\eta,{\bf {\cal E}})$ by $E_1({\bf R},\eta,{\bf
{\cal E}})$. The next step is to perform a constrained minimization, computing
\begin{equation}
U({\bf Q})=\min_{{\bf Q}({\bf R},\eta)= {\bf Q}}
E({\bf R},\eta,0)
\label{eq:U0Q}
\end{equation}
where the constraint is not on the polarization, but on 
${\bf Q}({\bf R},\eta)$ where ${\bf Q}({\bf R},\eta)$ is the Ti
displacement relative to the average position of the other atoms in the
unit cell. ${\bf R}({\bf Q})$ and $\eta({\bf Q})$ will be defined as the values of the atomic coordinates and strain at the minimum.

The equilibrium energy $E_{\rm FC}({\bf {\cal E}})$, structure ${\bf R}({\bf
Q}_{\rm min}), \eta({\bf Q}_{\rm min})$ and polarization ${\bf P}_{\rm FC}({\cal
E})={\bf P}({\bf R}({\bf Q}_{\rm min}),\eta({\bf Q}_{\rm min}),0)$ are then
obtained by the minimization
\begin{equation}
E_{\rm FC}({\bf {\cal E}})=\min_{\bf Q}\{U({\bf Q}) - {\bf {\cal
E}}\cdot{\bf P}({\bf R}({\bf Q}),\eta({\bf Q}),0)\} \, .
\label{eq:minQ}
\end{equation}
However, the results will in general not be equal to those obtained with 
our $i=1$ expression
\begin{equation}
E({\bf {\cal E}})=\min_{\bf P}
\{\min_{{\bf P}(\bf R,\eta,0)={\bf P}}{E({\bf R},\eta,0) - 
{\bf {\cal E}}\cdot{\bf P}\}} \, .
\label{eq:E1Fu}
\end{equation}
The reason is that
\begin{eqnarray}
E({\bf {\cal E}})
&=&\min_{\bf
R,\eta}\{{E({\bf R},\eta,0) - {\bf {\cal E}}\cdot{\bf P}({\bf R},\eta,0)\}}\nonumber\\
&=&\min_{{\bf Q}}\{\min_{{\bf Q}({\bf R},\eta)= {\bf Q}}
\{E({\bf R},\eta,0)
- {\bf \cal E}\cdot {\bf P}({\bf R},\eta,0)\}\}\nonumber\\
&\leq& \min_{{\bf Q}}
\{\left.\{E({\bf R},\eta,0)- {\bf {\cal E}}\cdot {\bf P}({\bf R},\eta,0)\}
\right\vert_{{\bf R}({\bf Q}),\eta({\bf Q})} \} \nonumber\\
&=&E_{\rm FC}({\bf {\cal E}}).
\label{eq:E2Fu}
\end{eqnarray}
The point is that the coordinates ${\bf R}({\bf Q})$ and $\eta({\bf Q})$ that
minimize $E$ alone at fixed ${\bf Q}$ are generally not the same as those
that would minimize the combination $(E-{\cal E} \cdot{\bf P})$ at fixed ${\bf Q}$.
At nonzero $\cal E$, equality will only be obtained under very special 
circumstances, for example, if the surfaces of constant ${\bf P}$ in
${\bf R}, \eta$ space coincide with the surfaces of constant ${\bf Q}$, 
at least for the relatively low energy structures. More specifically, 
the polarization should be a function only of the Ti displacement 
relative to all other atoms independent of the detailed arrangement 
of those atoms and of the strain. This is not unreasonable for small
enough fields in BaTiO$_3$, in which the soft mode is 
almost a pure Ti displacement and is well isolated in energy 
from other polar modes. However, as we will see in the following 
discussion, this proportionality is never quite exact even for small 
fields, and the discrepancies grow rapidly as the fields get larger.

%========================================================================
\section{Methodology}
\label{sec:Method}
%========================================================================
\subsection{Minimization procedure}
\label{sec:Method-Min}
%------------------------------------------------------------------------
We now describe in detail how the minimization of
$E({\bf R},\eta,0)$ respecting the constraint $P({\bf R},\eta,0) =
{\bf P}$ is implemented. Eq.~(\ref{eq:F1}) shows that this
constraint can be imposed by
a Lagrange multiplier $\lambda$. Therefore, we need to solve
the stationary-value problem described by Eqs.~(\ref{eq:mincond}).

Suppose we make a trial guess of the initial coordinates ${\bf R}_0$ 
and strains $\eta_0$ for the desired structure.
The energy $E({\bf R},\eta,0)$ can be expanded up
to second order in $\delta{\bf R}={\bf R}-{\bf R_0}$ and 
$\delta\eta=\eta - \eta_0$ as

\begin{eqnarray}
E({\bf R},\eta,0) &=& E({\bf R}_0,\eta_0,0)
+\sum_{i\alpha}{(-F_{i\alpha})\delta R_{i\alpha}} \nonumber\\
&+& \sum_{\mu\nu}{(-\sigma_{\mu})\delta\eta_{\nu}}
+\frac{1}{2}\sum_{\alpha\beta,ij}{K_{\alpha\beta}^{ij}
\delta R_{i\alpha}\delta
R_{j\beta}}\nonumber\\
&+&\frac{1}{2}\sum_{\mu\nu}{c_{\mu\nu}\delta\eta_\mu\delta\eta_\nu}
+ \sum_{i\alpha,\mu}\gamma^\mu_{i\alpha}\delta R_{i\alpha}\delta\eta_\mu
\end{eqnarray}
where $F_{i\alpha}$  
are the Hellmann-Feynman forces, $\sigma_{\mu}$
are the stresses in Voigt notation, 
$K_{\alpha\beta}^{ij}$ are the force-constant matrix elements, 
$c_{\mu\nu}$ are the elastic constants, and
$\gamma_{i\alpha \mu}$ are the coupling parameters between the internal
coordinates and strains. The corresponding variation in the polarization
${\bf P}({\bf R},\eta,0)$ is
\begin{eqnarray}
{\bf P}_\alpha({\bf R}, \eta,0) &=& {\bf P}_\alpha({{\bf R}_0, \eta_0},0)+
\sum_{j\alpha}{Z^i_{\alpha\beta} \delta R_{i\beta}} 
+ \sum_{\mu}{e_{\alpha\mu}\delta\eta_\mu}
\label{eq:P0exp}
\end{eqnarray}
where $Z^i_{\alpha\beta}=\partial P_\alpha/\partial R_{i\beta}$
and $e_{\alpha\mu}=\partial P_\alpha/\partial\eta_\mu$ are respectively 
the dynamic effective charge and piezoelectric tensors.

Eqs.~(\ref{eq:mincond}) lead to the linear system of equations
\begin{equation}
\left( \begin{array}{ccc}
K        &\gamma &Z^* \\
\gamma   &c      &e   \\
Z^*      &e      &0
\end{array}\right) \left(\begin{array}{c}  \delta R \\ \delta\eta\\
\lambda \end{array}\right) = \left(\begin{array}{c} F \\ \sigma
\\-{\mathcal P}
\end{array}\right)
\label{eq:linear-solve}
\end{equation}
for $\delta R$, $\delta\eta$ and $\lambda$, 
where ${\mathcal P}$ on the right-hand side  denotes the
difference between the initial and target values of ${\bf P}$. At
each step of the minimization, we compute $\delta{\bf R}$ and
$\delta\eta$, and obtain the new coordinates and strains via
${\bf R_0^{\rm new}}={\bf R}_0+\delta{\bf R}$ and $\eta_0^{\rm
new}=\eta_0+\delta\eta$. 
Then ${\bf R_{\rm new}}$ and $\eta_{\rm new}$ are chosen as the
new ``trial'' coordinates and strains. This is repeated until
convergence is achieved.

For a practical implementation of this procedure we use
density-functional perturbation theory, which allows us 
to compute the coefficients 
$K_{\alpha\beta}^{ij}$ and $Z^i_{\alpha\beta}$ efficiently. The forces $F$
and the stresses $\sigma$ are calculated by the Hellmann-Feynman theorem with Pulay
corrections\cite{Froyen-Cohen} for the stresses.  
However, the computation of the remaining quantities in
Eq.~(\ref{eq:linear-solve}),
involving derivatives with respect to strain, requires two additional
comments.
First, the DFPT calculation of $\gamma$, $c$ and $e$ is not 
yet
implemented in the current version of the {\tt ABINIT} package
(see Sec.~\ref{sec:Method-Details}) and 
the finite-difference calculation of these
quantities would be exceedingly tedious.
However, we will show that an alternative indirect minimization can be carried
out by means of a Devonshire-type Hamiltonian.\cite{Devonshire}
Details will be given
in Section~\ref{sec:Dielectric} and \ref{sec:SPT}.
Second, the most efficient way to compute $\cal P$ is with
a discretized Berry-phase expression. However, the dependence
on ${\bf R}$ and $\eta$ of the resulting polarization  
corresponds to the DFPT derivatives exactly only in the limit
of a dense k-point sampling mesh. This issue is discussed
and resolved in detail in the next subsection.

Before concluding this subsection, we note that the higher-order
formalism can be implemented in an analogous way. However, 
additional energy derivatives would be needed.
The details of the treatment for $i=2$
are presented in Appendix~\ref{app:higherorder}.
For the following study, we will restrict ourselves to 
the first-order case described by Eq.~(\ref{eq:E1}-\ref{eq:F1}).

%------------------------------------------------------------------------
\subsection{DFPT computation of derivatives of the discretized Berry-phase polarization
}
\label{sec:Method-compatibility}
%-----------------------------------------------------------------------
In the implementation of the minimization procedure
(Eq.~\ref{eq:linear-solve}),
a practical
problem arises in connection with the calculation of the dynamical
effective charges $Z^*$ and polarization ${\bf P}$. By definition, they should be
related by 
\begin{equation}
Z^i_{\alpha\beta} = V\frac{\partial P_\alpha}{\partial R_{i\beta}} \, ,
\label{eq:Zdef}
\end{equation}
where $\alpha$ and $\beta$ are Cartesian directions, $i$ is the index for the
atom, and $V$ represents the unit cell volume.

However, when the discretized Berry-phase expression is used to compute
${\bf P}$ and the DFPT expression is used to compute $Z^*$ on the same
{\bf k}-point mesh,
Eq.~(\ref{eq:Zdef}) is not
satisfied exactly.  The discrepancy vanishes in the limit of a dense
{\bf k}-point mesh, but in a practical calculation,
which must use a finite mesh, it will result in an inconsistency in the equations
for the minimization.

In the Berry-phase theory\cite{KVBerry} the polarization is
\begin{equation}
P^{\rm BP}_{\alpha}= \frac{ife}{(2\pi)^3}\int_{\rm
BZ}\sum^{\rm occ}_{m}{\left\langle u_{m{\bf k}}
\left\vert\frac{d}{dk_\alpha}\right\vert u_{m{\bf
k}}\right\rangle}d{\bf k} \, .
\label{eq:Pdef}
\end{equation}
where $f=2$ for spin degeneracy.  ${\bf P}^{\rm BP}$ is computed in
practice using a discretized formula which, for the case of 
isolated bands, takes the form 
\begin{equation}
P_\alpha= -\frac{fe} {(2\pi)^3} \int_A{\rm dk_{\perp}}
\sum^{\rm occ}_{m} {\rm Im}\ln\prod_{{\bf k}\in{\cal S}({\rm k}_{\perp})}
{\langle u_{m{\bf k}}\vert u_{m,{\bf k}+{\bf b}}}\rangle
\label{eq:P-berry}
\end{equation}
where the integration over the 2D ${\rm k_\perp}$ plane perpendicular to
direction $\alpha$ is replaced in practice by a summation over
a 2D mesh. The product runs over a string
${\cal S}({\rm k}_{\perp})$ of $\bf k$-points running parallel to
direction $\alpha$ at a given ${\rm k_\perp}$. $\bf b$ is
the separation between neighboring points along the string and $f=2$ is 
the spin degeneracy factor.   
The composite-band formulation is presented in Appendix~\ref{app:multi}.

In DFPT,\cite{GonzeLee} there are three equivalent expressions for the $Z^*$ tensor.
The first is the change in the polarization due to the first-order change
in the wavefunctions resulting from an atomic displacement:
\begin{equation}
Z^{i}_{\alpha\beta}=
  \frac{ifeV}{(2\pi)^3}
  \int_{\rm BZ} \sum^{\rm occ}_{m}
  \left\langle
    \frac{\partial u_{m{\bf k}}}{\partial R_{i\alpha}}
  \left\vert
    \frac{\partial u_{m{\bf k}}}{\partial k_\beta}
  \right.
  \right\rangle \, d{\bf k}\, ,
\label{eq:Z1-DFPT}
\end{equation}
where $\partial u_{m{\bf k}}/\partial R_{i\alpha}$
is the first-order change of the 
wavefunction due to the perturbation by displacing an atom belonging to 
the  $i$th
sublattice along the $\alpha$ axis. Alternatively, Z$^*$ can be computed as the
derivative of the force along direction $\alpha$ on an atom in the 
$i$th sublattice  with respect to an electric field along direction $\beta$,
\begin{eqnarray}
Z^{i}_{\alpha\beta}&=& 2\left[\frac{V}{(2\pi)^3}
\int_{\rm BZ}\sum^{\rm occ}_{m}
f{\left\langle u_{m{\bf k}}
\left\vert\frac{\partial v_{\rm ext}}
{\partial R_{i\alpha}}\right\vert\frac{\partial u_{m{\bf
k}}}{\partial{\bf\cal E}_\beta}\right\rangle d{\bf k}}\right. \nonumber\\
&+& \left.\frac{1}{2}\int_{V}\frac{\partial v_{\rm xc}({\bf r})}{\partial
R_{i\alpha}}\frac{\partial n({\bf r})}{\partial{\bf{\cal E}_\beta}}d{\bf
r}\right]  \, ,
\label{eq:Z2-DFPT} 
\end{eqnarray}
where $\partial u_{m{\bf k}}/\partial{\bf\cal E}_\beta$ 
is the first-order change of the wavefunctions due to the electric field, and 
$\partial v_{{\rm ext}}/\partial R_{i\alpha}$
and $\partial v_{\rm xc}({\bf r})/\partial R_{i\alpha}$
are, respectively, the first-order derivatives of
the external potential and the exchange-correlation potential with
respect to a ${\bf q}=0$ displacement.\cite{expla-v}
The third expression (omitted here) includes both types of first-order wavefunction
changes, and is stationary with respect to small errors in the
first-order wavefunctions.

In DFPT, first-order changes $\vert\psi^{(1)}\rangle$ in
the wavefunctions with respect to a perturbation can be
computed as self-consistent solutions of 
the first-order Sternheimer equations\cite{BGT}
\begin{equation}
P_c(H-\epsilon_m)P_c\vert\psi_m^{(1)}\rangle = -P_c H^{(1)}\vert\psi_m^{(0)}\rangle
\label{eq:pert}
\end{equation} 
subject to a ``parallel transport'' gauge constraint\cite{Gonze95}
\begin{equation}
\langle\psi_n^{(0)}\vert\psi_m^{(1)}\rangle = 0 \, ,
\label{eq:gauge}
\end{equation}
where $H^{(1)}$ is the first-order change in $H$ and $P_c$ is the 
projection operator onto the subspace of the
conduction bands, and $n$ and $m$ run only over the valence bands. 

In the case of the electric-field perturbation with field 
in Cartesian direction $\alpha$, the Sternheimer equation takes the
form\cite{BGT,Gonze97}
\begin{equation}
P_c(H-\epsilon_{m{\bf k}})P_c\left\vert \frac{\partial u_{m{\bf
k}}}{\partial{\cal E}_\alpha}\right\rangle = -P_c H^{(1)} \vert
u_{m{\bf k}}\rangle \, .
\label{eq:field-Stern}
\end{equation}
where 
\begin{equation}
H^{(1)}= -i\frac{\partial}{\partial k_\alpha} + \frac{d v_{\rm H}}{d
{\cal E}_{\alpha}}+
\frac{d v_{\rm xc}({\bf r})}{d{\cal E}_\alpha} \, .
\label{eq:H1-field}
\end{equation}
As input to this equation, we need the quantity
$\partial u_{m{\bf k}}/\partial k_\alpha$ appearing on the right-hand
side.  This is obtained by solving a second Sternheimer
equation 
\begin{equation}
P_c(H-\epsilon_{m{\bf k}})P_c\left\vert \frac{\partial u_{m{\bf
k}}}{\partial k_\alpha}\right\rangle = -P_c \left(\frac{\partial H_{\bf
k}}{\partial k_\alpha}\right) \vert u_{m{\bf k}}\rangle \, ,
\label{eq:k-Stern}
\end{equation}
where $H_{\bf k} = \frac{1}{2}(-i{\bf\nabla}+{\bf k})^2 + v_{\rm KS}$ and thus
$\partial H_{\bf k}/\partial k_\alpha = -i\nabla_\alpha+k_\alpha$.
The presence of the operator $i{\partial/\partial k_\alpha}$
in Eq.~(\ref{eq:H1-field}) is a unique feature that arises from the
coupling ${{\bf {\cal E}}}\cdot{\bf P}$ 
between the macroscopic electric field and the polarization. 

Now we come to the main point of this subsection, which is that the
derivatives of $\bf P$ calculated from DFPT are not exactly the
derivatives of the discretized Berry-phase expression for $\bf P$
in practical calculations.
In particular, for a {\it given} {\bf k}-point sampling, the $Z^*$
computed by solving for $\partial \vert u_{m{\bf k}}\rangle/\partial
k_\alpha$ from Eq.~(\ref{eq:k-Stern}), and then computing $Z^*$ via
Eq.~(\ref{eq:Z1-DFPT}) or via Eqs.~(\ref{eq:field-Stern}) and
(\ref{eq:Z2-DFPT}), is {\it not} exactly equal to the $Z^*$
computed from finite differences of the Berry-phase polarization
in Eq.~(\ref{eq:P-berry}). The numerical 
discrepancy between the Berry-phase
polarization and the $Z^*$ computed from the DFPT affects 
the application of the constrained minimization scheme proposed
in Section~\ref{sec:Method-Min} because it introduces an inconsistency into
the linear system in Eq.~(\ref{eq:linear-solve}).

Our cure for this problem is to modify the algorithm by which
$\partial\vert u_{m{\bf k}}\rangle/\partial k_\alpha$ is calculated
within the DFPT framework. Instead of solving the Sternheimer equation
(\ref{eq:k-Stern}), 
we calculate $\partial\vert u_{m{\bf k}}\rangle/\partial
k_\alpha$ from finite differences of the ground state Bloch
wavefunctions $\vert u_{m\bf k}\rangle$ between the neighboring ${\bf
k}$-points along the $\alpha$-direction.  This approach corresponds
to the ``perturbation expansion after discretization'' formalism
discussed by Nunes and Gonze in Ref.~\onlinecite{NG}.

We illustrate our approach here for the case of a single band in
one dimension. The generalization to the three-dimensional
multi-band case is postponed to Appendix~\ref{app:multi}.

For a single band in one
dimension, the Berry-phase polarization is
\begin{equation}
P = -\frac{fe}{2\pi}\sum_k {\rm Im} \ln \langle u_k \vert u_{k+b} \rangle
\label{eq:berry-1d}
\end{equation}
and its first-order variation reflecting a first-order change in
the wavefunction $\vert\delta u_{k}\rangle$ is
\begin{eqnarray}
\delta P &=& -\frac{fe}{2\pi}\sum_k {\rm Im}  \left[\frac{\langle\delta u_k\vert
u_{k+b}\rangle}{\langle u_k\vert u_{k+b}\rangle}+ \frac{\langle u_k\vert\delta
u_{k+b}\rangle}{\langle u_k\vert u_{k+b}\rangle}\right] \nonumber \\
&=&  -\frac{fe}{2\pi}\sum_k {\rm Im} \left[\frac{\langle\delta u_k\vert
u_{k+b}\rangle}{\langle u_k\vert u_{k+b}\rangle}- \frac{\langle \delta
u_k\vert u_{k-b}\rangle}{\langle u_k\vert u_{k-b}\rangle}\right] \nonumber \\
&=& \frac{feb}{\pi}\sum_k {\rm Re}\langle\delta u_k\vert v_k\rangle 
\label{eq:dP_finite}
\end{eqnarray}
where 
\begin{equation}
\vert v_k\rangle =
  \frac{i}{2b}\left[\frac{\vert u_{k+b}\rangle}{\langle u_k\vert
u_{k+b}\rangle} - \frac{\vert u_{k-b}\rangle}{\langle u_k\vert
u_{k-b}\rangle} \right]
\label{eq:vk-finite}
\end{equation}
is understood to be a finite-difference approximation to
$i\partial \vert u_k\rangle/\partial k$.  Note that
Eqs.~(\ref{eq:dP_finite}-\ref{eq:vk-finite}) are manifestly
gauge-independent in the sense of being independent of the choice of
phases for the $\vert u_k\rangle$.  

In the three-dimensional multi-band
case, we just need to replace $v_k$ of
Eq.~(\ref{eq:vk-finite}) by its generalization $v_{m{\bf k},\alpha}$
representing $\partial u_{m\bf k}/\partial k_\alpha$ as discussed in
Appendix B; $v_{m{\bf k},\alpha}$ is gauge-independent in the more general
sense of being invariant with respect to a unitary rotation among occupied
bands on neighboring $\bf k$-points.  This $v_{m{\bf k},\alpha}$
can then be substituted for $\partial u_{m\bf k}/\partial k_\alpha$ in 
Eq.~(\ref{eq:Z1-DFPT}) to compute $Z^*$. Or equivalently, it can
be inserted into Eq.~(\ref{eq:field-Stern}) to compute
$\partial u_{m\bf k}/\partial{\cal E}_\alpha$, which in turn can
be substituted into Eq.~(\ref{eq:Z2-DFPT}) to compute $Z^*$.
In either case, we are guaranteed to obtain the same values of $Z^*$
as would be derived from a series of finite-difference calculations
of polarization vs.~atomic displacement using the same $\bf k$-point
set.  This is because
Eqs.~(\ref{eq:dP_finite}-\ref{eq:vk-finite}) are derived directly
from the Berry-phase polarization expression of Eq.~(\ref{eq:Pdef})
using the same ${\bf k}$ mesh.  Moreover, because the Berry-phase
polarization (including ionic contributions) is independent of
origin, it also follows that the acoustic sum rule\cite{Pick}
$\sum_{i}{Z^{i}_{\alpha\beta}}=\delta_{\alpha\beta}$ on the
components of the dynamic effective charges will be satisfied
exactly, which is not the case in conventional linear-response
calculations of $Z^*$.

%----------------------------------------------------------------------
\subsection{Computational Details}
\label{sec:Method-Details}
%---------------------------------------------------------------------- 
We carried out all the {\it ab initio} calculations using the {\tt ABINIT} 
package,\cite{expla-AB}
in which we have implemented the above algorithm.
{\tt ABINIT} uses a plane-wave basis and 
provides multiple norm-conserving (NC) and extended NC pseudopotentials. 
The discretized formula for the wavefunction derivatives with respect
to the wavevectors, Eq.~(\ref{eq:vnkmul}), is introduced in a new
subroutine {\tt dudk.f}, a key ingredient that allows us
to carry out the constrained-polarization minimization scheme.  
  
In Sec.~\ref{sec:ISB}, in order to construct the pseudopotentials 
for the virtual atoms\cite{Sai} that enter the heterovalent system
Ba(Ti$-\delta$,Ti,Ti$+\delta$)O$_3$, 
we utilize the FHI atomic code\cite{FHI} that
generates Troullier-Martin separable norm-conserving
pseudopotentials.\cite{TM} However, the FHI pseudopotential generation
scheme only allows one projector within each angular momentum channel, thus
preventing the inclusion of the $3s$ and $3p$ states, in addition to
$3d$ and $4s$ states, in the valence for the Ti 
pseudopotential. (The same problem occurs for the $5s$ and $5p$ states
for the Ba atom.) We generate the pseudopotential in ionized
configurations $3s^23p^63d^24s^0$ for Ti and $5s^25p^66s^0$ for Ba. 
We used the exchange-correlation energy functional in the
Ceperley-Alder\cite{Ceperley-Alder} form with 
Perdew and Wang\cite{Perdew-Wang} parameterization.
 
The studies described in Sections~\ref{sec:BT} to \ref{sec:SPT} have been
performed with the highly transferable extended norm-conserving
pseudopotentials proposed by Teter.\cite{Teter} A Perdew-Zunger\cite{Perdew-Zunger}
parameterized Ceperley-Alder exchange-correlation functional was
used. These pseudopotentials include the Pb $5d$, $6s$ and $6p$,
the Ba $5s$, $5p$ and $6s$, the Ti $3s$, $3p$, $4d$ and $4s$, and the O
$2s$ and $2p$ electrons in the valence states.  
 
We have used an energy cutoff of 35 Ha throughout. The
integrals over the Brillouin zone have been replaced by a sum over
a $4\times 4\times 4$ ${\bf k}$-point mesh. Both the ${\bf k}$-point sampling and
the energy cutoff have been tested for good convergence of the phonon
eigenvalue and eigenvector properties. We use the same ${\bf k}$-point mesh
for the Berry-phase calculations. 
Convergence of the relaxations requires the 
Hellmann-Feynman forces to be less than 0.02\,eV/\AA. 
(In the constrained minimization procedure
described by Eq.~(\ref{eq:linear-solve}), 
the forces that are tested for convergence 
are the ones after projection onto the constant-${\bf P}$ subspace.)

%========================================================================
\section{Sample Applications}
\label{sec:Appli}
%========================================================================
In this section, we illustrate the theory within the first-order ($i=1$)
formalism (see Section~\ref{sec:forma}) by applying it to a series of
problems involving ferroelectric, dielectric and piezoelectric properties. In 
particular, we emphasize that the main purpose of these calculations is to 
exhibit and understand the nonlinearity in the structural
response of the ferroelectric systems to an electric 
field. Such studies have not previously been widely pursued. 
We have used BaTiO$_3$, PbTiO$_3$ and a short-period superlattice
structure as our example systems. 

%--------------------------------------------------------------------------
\subsection{Inversion symmetry-breaking system} 
\label{sec:ISB}
%--------------------------------------------------------------------------
In a conventional ABO$_3$ perovskite such as BaTiO$_3$, the cubic
symmetry of the high-temperature paraelectric phase is spontaneously
broken at the transition to the ferroelectric phase.  The atomic
displacements that occur in the ferroelectric phase give rise to an
associated lattice strain, and the ferroelectric state is characterized
by a switchable polarization because of the occurrence of degenerate
energy minima that are connected by the broken symmetry operations.

Recently, using DFT total-energy methods, we (Sai, Meyer and
Vanderbilt\cite{Sai}) studied a new class of cubic perovskite compounds in
which the composition is modulated in a cyclic sequence of three
layers on the A site (i.e., $(AA'A'')BO_3$ structures) or on the B
site (i.e., $A(BB'B'')O_3$ structures). The inversion
symmetry that was present in the high-symmetry cubic structure is
now permanently broken in these materials by the alternating
compositions ordered along the lattice growth direction. This gives
rise to important qualitative differences in the energetic behavior
of these compounds relative to simple ABO$_3$ perovskites.  Most
interestingly, it was shown that by using heterovalent
compositional substitutions, the strength of the breaking of the
inversion symmetry could be tuned through an enormous range,
suggesting that such systems could be very promising candidates for
new materials with large piezoelectric and other dielectric
response properties.\cite{Sai}

\begin{figure}
\centerline{\epsfig{file=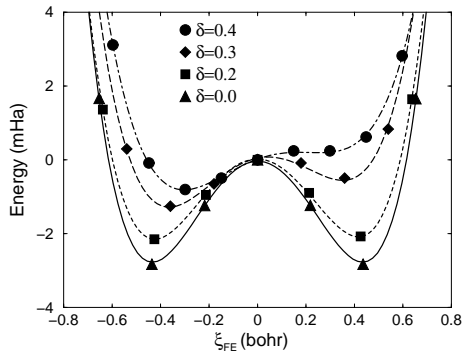,angle=270,width=6cm}}
\caption{Energy {\it vs}.\ displacement along the line connecting
the two energy minima in the Ba(Ti-$\delta$,Ti,Ti+$\delta$)O$_3$ system,
plotted for several values of $\delta$.}
\label{fig:ISB_Eu}
\end{figure}

Such compositionally modulated structures were studied within a
model system Ba(Ti$-\delta$,Ti,Ti$+\delta$)O$_3$ where the two
atomic species that alternate with Ti on the $B$ site are virtual
atoms that we constructed by varying the nuclear charge of Ti
by $\pm\delta$.
Therefore, as $\delta$ is tuned
continuously from 0 to 1, we can simulate a set of systems
evolving from a conventional BaTiO$_3$ ferroelectric system
to a heterovalent Ba(Sc$_{1/3}$Ti$_{1/3}$V$_{1/3}$)O$_3$ one
in which all three alternated species are from neighboring
columns in the periodic table.

As a consequence of the
compositionally broken inversion symmetry, the thermodynamic
potential associated with the FE instability does not have the
usual symmetric double-well form.  Instead, it takes the form of
an asymmetric double well, or even of an asymmetric single well,
depending on the strength of the compositional perturbation that
breaks the symmetry.  In normal ferroelectrics, it is sufficient
to locate one minimum of the double-well potential; the other is
then obviously given by applying the inversion operation. Here,
this no longer works.  A steepest-descent minimization starting
from the ideal structure typically arrives at the primary (deeper)
minimum, but the secondary (shallower) minimum can be rather
difficult to find in practice.

\begin{figure}
\centerline{\epsfig{file=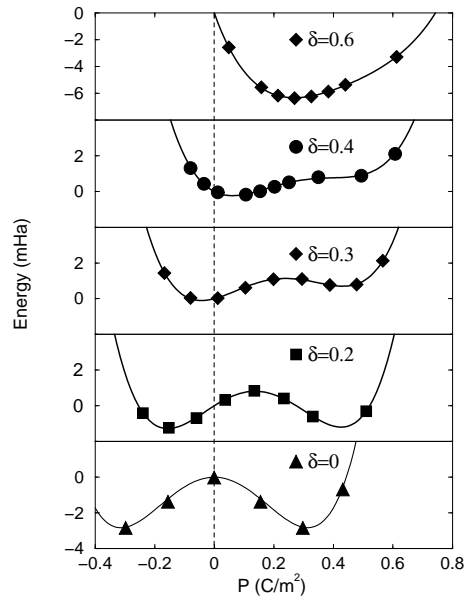,width=6cm}}
\caption{Energy {\it vs} polarization $P$ in
Ba(Ti-$\delta$,Ti,Ti+$\delta$)O$_3$ at different $\delta$. Note the
saddle point of $E(P)$ shifts in the direction of the secondary
(shallower) minimum as $\delta$ increases.}
\label{fig:ISB_EP}
\end{figure}

A procedure was described in Ref.~\onlinecite{Sai} that allows
one to search for both minima when they coexist.  To illustrate the
procedure, we plot in Fig.~\ref{fig:ISB_Eu}, for several values
of $\delta$, the energy as a function of displacement amplitude along
the straight line in the 15-dimensional parameter space connecting
the primary and secondary minima. (The direction along this line
is taken to define the ``FE direction'' with $\xi_{\rm FE}=0$ being the
midpoint between minima.)  Unfortunately, however,
it is not possible to plot such a curve for $\delta>0.4$, since only
a single minimum exists in this range of $\delta$.

Here, we demonstrate how the current method allows for a much more
natural treatment of these systems, especially at large $\delta$.
At a fixed value of polarization, we calculate $F({\bf P})$
as in Eq.~(\ref{eq:F1}).  That is, we minimize the total energy over
the internal coordinates subject to the constraint that the spontaneous
polarization has a fixed value, following the procedure described in
Section~\ref{sec:Method}.  As in our previous work, this
is done in a fixed tetragonal cell.

The energy as a function of the polarization  for several values of
$\delta$ is illustrated in Fig.~\ref{fig:ISB_EP}. We obtain
a similar energy evolution as in Fig.~\ref{fig:ISB_Eu}. However, there
are two important qualitative differences.  First, the new procedure
is not limited to the range of $\delta$ in which both minima exist.
At larger $\delta$ (e.g., $\delta=0.6$), where the secondary minimum
has disappeared due to a strong symmetry-breaking perturbation, the new
procedure allows the mapping of the energy to be carried out just as easily
as at smaller $\delta$.  Second, the horizontal axis of the figure now
has a physical meaning of polarization.  For example, a glance at
Fig.~\ref{fig:ISB_EP} shows an interesting feature, namely that
the saddle point is also polarized, unlike in a normal ABO$_3$ compound.

\begin{figure}
\centerline{\epsfig{file=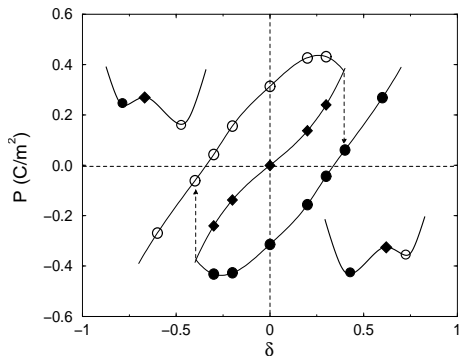,width=6cm}}
\caption{Calculated polarization at the left minimum (solid circle),
right minimum (open circle), and saddle point (diamond) in
Ba(Ti-$\delta$,Ti,Ti+$\delta$)O$_3$.  Left (right) minimum
is the principal one for $\delta>0$ ($\delta<0$), as shown in
insets.}
\label{fig:ISB_Pmin}
\end{figure}

We investigated this interesting feature further by plotting in
Fig.~\ref{fig:ISB_Pmin} the polarization at the saddle point, as well
as at the energy minima, as a function of $\delta$.  In the ``normal''
case $\delta=0$, the saddle point is unpolarized and the two
equivalent minima carry equal and opposite polarizations. However,
all the stationary points are seen to be shifted in the direction 
of the shallower minimum as $\delta$ is turned on.  As a
consequence of these shifts, the polarization of the primary
minimum changes sign near $\delta\simeq0.4$.  At the critical
$\delta$ the saddle point and secondary minimum meet and annihilate.
Returning to small values of $\delta$, a closer analysis (not
shown) indicates that the polarization at the saddle point
increases in proportion to $\delta^3$.
This observation agrees well with previous studies\cite{Sai}
showing that certain other measures of the effect of the symmetry-breaking
perturbation also scale as $\delta^3$.

In summary, we have illustrated the convenience and power of our new
method in the context of recent work on a new class of ferroelectric
materials with compositionally broken inversion symmetry.  The new
approach is especially useful for studying the case where the compositional
perturbation is so strong that only a single local minimum survives.
In the former procedure, the definition of the FE direction was based
on the location of two local minima, and was therefore useless when one
of the minima had disappeared.  On the contrary, in the new method the
energy surface can be straightforwardly mapped out as a function of
polarization, regardless of whether the secondary minimum exists or not.
Moreover, expressing the behavior as a function of polarization provides
a much more informative picture of the system.  For example, certain
interesting and non-trivial behaviors of the polarizations at the
saddle points and minima can be elucidated.

%-----------------------------------------------------------------------
\subsection{Structural response in BaTiO$_3$}
\label{sec:BT}
%----------------------------------------------------------------------

In this section, we apply our approach to BaTiO$_3$, one of the
most-studied perovskite ferroelectric compounds. It undergoes a sequence 
of structural phase transitions with decreasing temperature: from the 
high-temperature cubic to
the tetragonal phase at $130^\circ$C, then to an orthorhombic phase at 
$5^\circ$C and finally to the ground-state rhombohedral structure at 
$-90^\circ$C. \cite{LG} The three 
successive phases, distortions of the cubic perovskite structure, are  
characterized by spontaneous polarizations
aligned along the [001], [011], and [111] directions respectively.  

\begin{figure}
\centerline{\epsfig{file=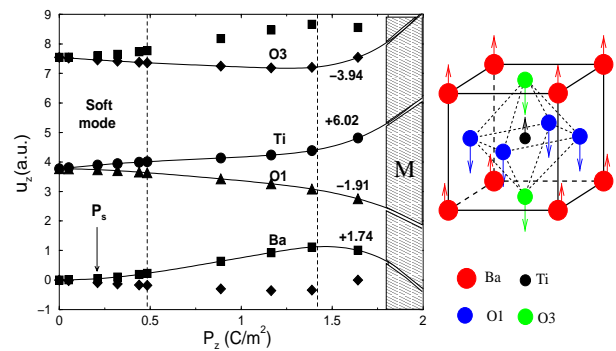,width=8cm}}
\caption{Fully relaxed $z$ coordinates for each atom in BaTiO$_3$
(see unit cell at right) as a function of
polarization $P_z$ in the simple-cubic lattice. Curves are cubic-spline
fits to calculated points; top- and bottom-most points
correspond to translational images of Ba and O3 atoms, respectively,
in neighboring unit cells.
$P_s$ marks the spontaneous polarization.  The
Born effective charge $Z^*$ for each atom is marked at $P_z=1.64{\rm
C/m}^2$. Shaded area indicates metallic regime.}
\label{fig:BTO}
\end{figure}

Here, we focus on the dependence of the internal structural
parameters of BaTiO$_3$ on the polarization ${\bf P}$. Our calculation
is restricted to allow only atomic displacements along the $\hat z$ direction in 
a fixed simple cubic lattice, with full relaxation of the internal structural
parameters at fixed polarization ${\bf P}=P_z \hat z$ to yield equilibrium coordinates
${\bf R}_{\rm eq}(P_z)$.
In this way, we can investigate the contribution of the internal structural 
parameters alone, decoupled from the strain degrees of freedom,
to the structural response to an electric field, providing a first step towards 
understanding the nonlinearities of the 
total structural response expected with increasing ${\cal E}$. 

Experimentally, BaTiO$_3$ is known to have a cubic lattice constant of
7.547 a.u.\cite{Landolt2} Our LDA calculation yields an
equilibrium lattice constant of 7.45 a.u., 1.3\% smaller than the
experimental value, an error typical of the LDA. As is well known, 
the ferroelectric instability depends
sensitively on the crystal volume.\cite{Cohen-Krakauer92,Cohen,KV94}
We therefore choose to work at the experimental cubic lattice constant.

The spontaneous polarization $P_s$ is obtained by full relaxation of the
internal structural parameters, and is found to be $0.21{\rm C}/{\rm m}^2$.
The relaxed internal coordinates for each Ba, Ti,
O1 and O3 atom are plotted as a function of $P_z$ in
Fig.~\ref{fig:BTO}. 
For $P_z>P_s$, the state can be realized as an
equilibrium state in an appropriate fixed electric field, while states with $P_z<P_s$ 
are local maxima of $F(P_z)-{\cal E} P_z$ for some value of $\cal E$.
For example, the value of $\cal E$ corresponding to $P_z=0.48{\rm C}/{\rm m}^2$
(approximately twice $P_s$) is 16 MV/cm. 

\begin{figure}
\centerline{\epsfig{file=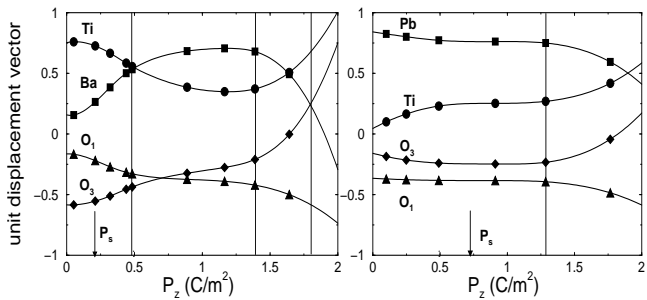,width=8.5cm}}
\caption{The component of the unit displacement vector $\xi_z$
corresponding to each atom in BaTiO$_3$ (left) and PbTiO$_3$
(right) as a function of polarization $P_z$.
In the left panel, vertical lines demarcate the regimes of
BaTiO$_3$ soft-mode-like, PbTiO$_3$ soft-mode-like, atom-pair, and
metallic behavior;
at right, the single vertical line separates soft-mode-like and
atom-pair regimes.}
\label{fig:xi}
\end{figure}

To focus on the dependence of the character of the distortion on 
the amplitude $P_z$, we define a ``unit displacement vector"
$(\xi^{\rm Ba}_z, \xi^{\rm Ti}_z, \xi^{\rm O1}_z, \xi^{\rm O3}_z)$ 
by normalizing the sum of the squared displacements to one.
At $P_z=P_s$, the unit displacement vector 
is found to be $(0.26, 0.73, -0.22, -0.55)$, 
closely resembling the unstable ferroelectric mode of cubic BaTiO$_3$
$(0.18,0.74,-0.18,-0.59)$ computed from a linear-response 
calculation. 
In Fig.~\ref{fig:xi}, we show the $P_z$ dependence of the components
of the unit displacement vector.
If the polarized state were obtained by freezing in a single polar mode, 
these components would be constant. 
The actual behavior is considerably more complicated.

Three distinct regimes for the atomic displacement pattern
can be clearly observed.
For $P_z$ below $\approx 0.48{\rm C}/{\rm m}^2$, the relative
displacements are similar in character to those of the soft mode. 
In this regime, the magnitudes of the Ba and O1 components increase with $P_z$,
while the magnitudes of the Ti and O3 components decrease.
For $P_z$ between roughly $0.48{\rm C}/{\rm m}^2$ 
and $1.4{\rm C}/{\rm m}^2$, the consequence of these
opposing trends is that the magnitudes of the Ba and O1 displacements
actually exceed those of Ti and O3, 
respectively, changing the character of the structural distortion.
At $P_z\approx1.4{\rm C}/{\rm m}^2$, 
the trend with $P_z$ reverses for Ba and Ti, so that as the polarization
increases further, the Ba and O1 atoms move together in a direction
opposite to that of Ti and O3.

\begin{figure}
\centerline{\epsfig{file=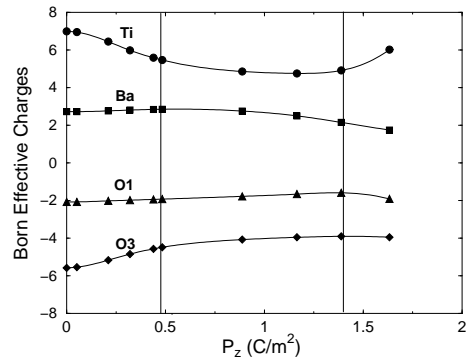,angle=270,width=6cm}}
\caption{Computed Born effective charges for each atom in
BaTiO$_3$ as a function of the polarization $P_z$.
For the cubic structure ($P_z$=0), we obtain
$Z^*_{\rm Ba}=2.72$, $Z^*_{\rm Ti}=6.99$, $Z^*_{\rm O_\perp}= -5.57$,
$Z^*_{\rm O_\parallel}= -2.07$.
}
\label{fig:BT_Zeff}
\end{figure}

The Born effective charges $Z^*$ are expected to be sensitive to
the internal structural parameters.  Figure \ref{fig:BT_Zeff} shows
the evolution of the computed $Z^*$ for each atom. 
Near the cubic structure, the
dependence of $Z^*_{\rm Ti}$ and $Z^*_{\rm O3}$ on $P_z$ is nearly quadratic, 
in agreement with previous calculations for BaTiO$_3$.\cite{Ghosez98} 
While $Z^*_{\rm Ba}$ and $Z^*_{\rm O1}$ are rather insensitive to $P_z$, 
the Born effective charge of Ti decreases by over 30\% from $P_z=0$ to
$1.4{\rm C}/{\rm m}^2$, with a corresponding increase in that of O3.
More specifically, $Z^*_{\rm Ti}$ drops to its
smallest value $+4.7$ while the magnitude of $Z^*_{\rm O3}$ is close to 
its smallest value $-3.9$.
This structural sensitivity can be understood as being related to the anomalous
values in the undistorted cubic perovskite structure,
which arise from the hybridization of Ti and O orbitals in the
Ti-O3 chains oriented along $\hat z$. As shown in
Fig.~\ref{fig:BTO}, when $P_z$ increases, the 
Ti are displaced towards one O3 neighbor and away from the other,
disrupting the chain and reducing the anomalous displacement-induced
current along the chain.\cite{Zhong-Coulomb}
Consequently, the magnitudes of the Born effective charges are reduced
towards the nominal valences +4 for Ti and $-$2 for O.

As the polarization and associated structural distortions become larger,
the bandstructure evolves correspondingly. 
We find significant changes in the
band structure for $P_z=1.8{\rm C}/{\rm m}^2$ relative to that of the
undistorted cubic perovskite structure. 
The hybridization between the Ti 3d and O 2p bands becomes more
significant, as expected from the decreased Ti-O3 distance.  
Some bands, such as the topmost O 2p band, lose the 
characteristic flatness that is usually seen in perovskites like
BaTiO$_3$\cite{KV94} and become much more dispersive.  Most
importantly, the band gap decreases with increasing $P_z$, extrapolating
to an insulator-metal transition just above $P_z=1.8{\rm C}/{\rm m}^2$.
The polarization is a meaningful quantity only in insulators, and therefore
calculation for higher values of $P_z$ cannot be considered.

We carried out an analogous calculation for PbTiO$_3$ in the 
cubic structure using the lattice
constant deduced from experiment, $a_0=7.5$ a.u, yielding a 
spontaneous polarization of $0.73{\rm C/m^2}$.
The unit displacement vector as a function of $P_z$ is shown in the right panel 
of Fig.~\ref{fig:xi}, where the pattern resembles that of BaTiO$_3$
at intermediate values of $P_z$. Thus, the pattern that is field-induced in
BaTiO$_3$ is characteristic of that of PbTiO$_3$ at zero electric field.
The high-$P_z$ regime sets in at around $1.5{\rm C/m^2}$,
with the Pb and O1 atoms moving together as a pair 
and Ti and O3 moving together as a second pair.

Some general observations can be made about the effects of an electric 
field on the internal structural parameters. At small fields, 
the cations and anions move independently, following the   
electrostatic force corresponding to the sign of the formal valence. 
However, once the field-induced distortions are large enough so that
short range interatomic repulsion prevents further compression of
Ba/Pb-O1 and Ti-O3 bonds, the further distortions acquire a new
character in which these atoms move as pairs.
From the computed values of the dynamic effective charges at
$P_z=1.64{\rm C}/{\rm m}^2$ (see Fig.~\ref{fig:BTO}), the 
Ba-O1 pair and Ti-O3 pair carry net charges very close to $-2$
and $+2$ respectively. 

In summary, we have shown that the ``simple" perovskite compound
BaTiO$_3$ exhibits significant nonlinearity in structure with
increasing polarization, corresponding to large electric fields,
while the atomic displacement pattern of PbTiO$_3$ is relatively
slowly varying. It is interesting to note that at large enough
polarization, 
the atomic displacement pattern of BaTiO$_3$ in fact
resembles that of PbTiO$_3$. 
This may help to shed some light on the factors responsible for the
differences in properties between alkaline earth and Pb-based
perovskites.

%--------------------------------------------------------------------
\subsection{Non-Linear Dielectric and Piezoelectric Response}               
\label{sec:Dielectric}
Tunability
of the dielectric and piezoelectric coefficients by an applied electric field,
a property of great technological importance, is expected to be especially large 
in ferroelectrics due to the dependence of these coefficients on
electric-field-induced structural changes such as
those reported for BaTiO$_3$ in the previous section.
This behavior can be quantified by the values of the nonlinear
dielectric and piezoelectric coefficients.
In this section, we formulate the calculation of these nonlinear coefficients
in our polarization-based framework, and give results for tetragonal
PbTiO$_3$.

The first step in this analysis is the computation of $F({\bf P})$
and $\eta({\bf P})$ from the minimization of $F({\bf R},\eta,{\bf
P})$ at fixed ${\bf P}$. This is followed by the minimization of
$F({\bf P}) - {\bf P}\cdot{\bf\cal E}$ at fixed $\cal E$, directly
yielding ${\bf P}({\bf\cal E})$ and $\eta({\bf\cal E})=\eta({\bf
P}({\bf\cal E}))$.
From the first derivative of ${\bf P}({\bf\cal E})$, we obtain the
field-dependent static dielectric susceptibility tensor
$\chi_{\alpha\beta}({\bf {\cal E}})$, with the nonlinear
coefficients defined through a small-$\cal E$ expansion
\begin{equation}
\chi_{\alpha\beta}({\bf {\cal E}}) =
\frac{1}{\epsilon_0}
\frac{\partial{{\bf P}_\alpha({\bf {\cal E}})}}{\partial{\bf
{\cal E}}_\beta}=\chi^{(1)}_{\alpha\beta}
+\chi^{(2)}_{\alpha\beta\gamma}{\cal E}_\gamma+{\cal O}({\cal E}^2) \, .
\label{eq:chi_field}
\end{equation} 
The relative dielectric tensor is given by
$\epsilon_{\alpha\beta}=\delta_{\alpha\beta}+\chi_{\alpha\beta}$.
Correspondingly, from the first derivative of $\eta({\bf\cal E})$,
we obtain the field-dependent piezoelectric tensor
$d_{\mu\beta}({\bf {\cal E}})$, with the nonlinear coefficients
defined through a small-$\cal E$ expansion
\begin{equation}
d_{\mu\beta}({\bf {\cal E}}) =
\frac{\partial{\eta_\mu({\bf {\cal E}})}}{\partial{\bf
{\cal E}}_\beta}=d^{(1)}_{\mu\beta}+d^{(2)}_{\mu\beta\gamma}{\cal E}_\gamma
+ {\cal O}({\cal E}^2) \, .
\label{eq:d_field}
\end{equation} 

In fact, in our present implementation we perform the minimization
of $F({\bf R},\eta,{\bf P})$ at fixed ${\bf P}$ in two separate
steps. First, we obtain a reduced free-energy density $F(\eta,{\bf
P})$ by minimizing with respect to ${\bf R}$ at fixed $\eta$ and
${\bf P}$. Further minimization with respect to $\eta$ to obtain
$F({\bf P})$ and $\eta({\bf P})$ allows the computation of the
zero-stress responses as in the previous paragraph. In addition,
this approach allows the computation of the clamped-strain
dielectric response, measured at frequencies above the resonant
frequency of the sample, through minimization of $F({\eta,\bf P}) -
{\bf P}\cdot{\bf\cal E}$ at fixed $\cal E$ and $\eta=\eta({\bf\cal
E})$, directly yielding $P({\eta({\bf\cal E}), \bf\cal E})$ and
$\chi({\eta({\bf\cal E}), \bf\cal E})$.  This two-step procedure is
also required, as mentioned in Section~\ref{sec:Method-Min}, by the
limitations imposed on the present implementation by the use of
ABINIT 3.1.  For present practical purposes, $F(\eta,{\bf P})$ is
obtained in a parameterized form by fitting a Landau-Devonshire
expansion to values of $F$ obtained by calculations for an
appropriate set of $\eta$ and ${\bf P}$.

We have applied this procedure to compute the nonlinear dielectric
and piezoelectric response of tetragonal PbTiO$_3$ to fields along
$\hat z$ using the $i=1$ expressions (Eq.~\ref{eq:F1}).
At this level of approximation, only lattice contributions to
$\chi_{\alpha\beta}({\bf {\cal E}})$ and $d_{\mu\beta}({\bf {\cal E}})$
are included, and their electric-field dependence arise only through
induced structural changes.  However, this is expected to be a good
approximation for PbTiO$_3$, where the lattice contribution to the
dielectric and piezoelectric responses dominates even at $T=0$.

With a field along ${\hat z}$, symmetry constrains the structural
response to consist of a tetragonal strain, specified by two independent
parameters $\eta_1=\eta_{xx}=\eta_{yy}$ and $\eta_3=\eta_{zz}$,
and a set of atomic displacements along the $\hat z$ direction described
by three
independent parameters. Correspondingly, we obtain an expression
for $F({\bf R},\eta,{\bf P})=F(\eta_1,\eta_3,P_z)$ by computing
$\min_{\bf R}{E({\bf R},\eta,{\cal E}=0)}$ for a set of tetragonal
cells with the constraint ${\bf P}({\bf R},\eta_1,\eta_3)=P_z\hat
z$.  The results are used to fit the parameters in a
Landau-Devonshire expression expanded around the minimum-energy
cubic structure ($a_0$=7.33 a.u.),
\begin{eqnarray}
F(\eta_1,\eta_3,P_z)&=& E_0
+\frac{1}{2}C_{11}(2\eta_1^2+\eta_3^2)+
C_{12}(2\eta_1\eta_3+\eta_1^2)\nonumber\\
&+&A_{200}P_z^2 + A_{400}P_z^4+A_{600}P_z^6\nonumber\\
&+&2B_{1yy}\eta_1 P_z^2+B_{1zz}\eta_3P_z^2 \, ,
\label{eq:1Dmodel}
\end{eqnarray}
where $P_z$ is the polarization per unit volume
and the truncations to sixth order in $P_z$ and to lowest order in
the elastic and polarization-strain coupling are found to be
sufficient within a standard least-squares fit.  The resulting
coefficients are shown in Table~\ref{tab:coeff1D}; statistical
analysis shows that the strain coupling parameters $B_{1yy}$ and
$B_{1zz}$ are the most sensitive to changes in the input
configuration energies.

\begin{table}
\caption{The values of the least-squares fitted
parameters in Eq.~(\ref{eq:1Dmodel}) at ${\cal E}=0$ in PbTiO$_3$. The
units are the appropriate combinations of Ha and$({\rm C/m}^2)^2$. }
\begin{center}
\begin{tabular} {cccc}
  Parameters & Values    & Parameters    & Values         \\\hline
  $E_0$      &$-$165.953        &$A_{400}$  &$\phantom{-}$0.005  \\
  $C_{11}$   &$\phantom{-}$4.374  &$A_{600}$  &$\phantom{-}$0.004 \\
  $C_{12}$   &$\phantom{-}$1.326  &$B_{1zz}$  &$-$0.199 \\
  $A_{200}$  &$-$0.003            &$B_{1yy}$  &$-$0.049
\end{tabular}
\end{center}
\label{tab:coeff1D}
\end{table}

We now use this expansion to compute the field dependence of the
strain and polarization under zero stress by minimizing
$F(\eta_1,\eta_3,P_z)-{\cal E}\cdot P_z$ with respect to $\eta_1$,
$\eta_3$ and $P_z$ to get $\eta_1({\cal E})$, $\eta_3({\cal E})$
and $P_z({\cal E})$. By first considering $\cal E$ = 0, we can
confirm the validity of the parameterization by comparing the
tetragonal structure obtained by minimizing the expression for
$F(\eta_1,\eta_3,P_z)$ given by Eq.~(\ref{eq:1Dmodel}) with
properties of the fully relaxed tetragonal ground-state structure
in zero electric field. In Table~\ref{tab:struc_tetra} we list the energy
difference $\Delta E$ between the tetragonal ground state and the
cubic structure, the spontaneous polarization $P_{z}({\cal E}=0)$,
and the lattice parameters, finding good agreement in all
respects.

\begin{table}
\caption{Comparison of the structural parameters computed by
minimizing Eq.~(\ref{eq:1Dmodel}) with those computed
from direct LDA calculation and those obtained from
experiment.\cite{Gavri,Haun}}
\begin{center}
\begin{tabular}{lcccc}
 & $\Delta E$ (mHa) & $P^{{\bf\cal E}=0}_{z}$ (C/m$^2$) & $a$
(Bohr) & $c$ (Bohr)\\\hline
 model& 0.86   & 0.67          & 7.324 & 7.487 \\
 LDA  & 0.90   & 0.65          & 7.310 & 7.484  \\
 exp  & -      & 0.75(295K)     & 7.373 & 7.852
\end{tabular}
\end{center}
\label{tab:struc_tetra}
\end{table}

Next, we consider nonzero $\cal E$. Minimizing first with respect
to $\eta$ gives a free energy
\begin{equation}
F(P_z)= A_{200} P_z^2 + \widetilde{A}_{400}P_z^4 + A_{600} P_z^6
\end{equation}
where
\begin{equation}
\widetilde{A}_{400}=A_{400}+
\frac{2c_{12}B_{1zz}B_{1yy}-c_{11}B_{1yy}^2-
\frac{1}{2}(c_{11}+c_{12})B_{1zz}^2}{(c_{11}+2c_{12})(c_{11}-c_{12})}\, ,
\end{equation}
and $\widetilde{A}_{400}$ is found to be 4.5$\times10^{-4}$HaC$^{-4}$m$^{-8}$.

\begin{figure}
\centerline{\epsfig{file=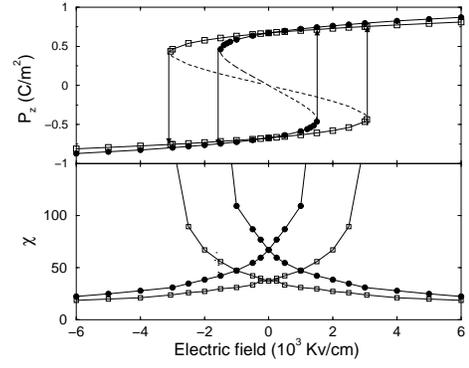,angle=270,width=6cm}}
\caption{Calculated polarization-vs.-electric-field
hysteresis loop (upper panel) and static susceptibility $\chi({\bf\cal
E})$ (lower panel) of PbTiO$_3$ under stress-free condition (solid circle)
and clamped-strain condition (open square).
Dashed line corresponds to the non-accessible state (saddle point in the
thermodynamic potential).}
\label{fig:PT_chi}
\end{figure}

Since $A_{200}<0$, $F(P_z)$ has a double well structure, so that
$F(P_z)-P_z{\cal E}$ has two local minima for small enough
values of $\cal E$.  The evolution of the two local minima with
${\cal E}$ can be summarized in the calculated hysteresis loop
shown in the upper panel of Fig.~\ref{fig:PT_chi}. We find an
intrinsic coercive field ${\bf\cal E}_{\rm c}$ of 1.5MV/cm.  From
Eq.~(\ref{eq:chi_field}), we can proceed to calculate the static
susceptibility $\chi_{33}({\bf\cal E})$ and the result is plotted
in the lower panel of Fig.~\ref{fig:PT_chi}. Fitting this to
Eq.~(\ref{eq:chi_field}), we find that the zero-field stress-free
susceptibility $\chi^{(1)}_{33}$ is $\chi^\sigma_{33}=67$, the
superscript $\sigma$ indicating stress-free conditions.

For the clamped-strain response at zero field, we fix $\eta$ at
$\eta({\cal E}=0)$. A different double well structure is obtained
for $F^{(\eta)}(P_z)$, resulting in a different hysteresis loop
shown in the same figure. We find an intrinsic coercive field
${\bf\cal E}_{\rm c}$ of 3MV/cm.  From fitting to
Eq.~(\ref{eq:chi_field}), we obtain $\chi^\eta_{33}=37$, the
superscript $\eta$ indicating the clamped-strain condition.
In Table~\ref{tab:dielectric}, these values
are compared with the reported experimental dielectric constants at both
the constant stress and clamped-strain condition\cite{Li} which were measured 
below and above the sample resonant frequencies respectively. 
The value for $\chi^\eta_{33}$ can also be compared 
with a previous first-principles
calculation.\cite{Waghmare}

\begin{table}
\caption{Comparison between theory and experiment\cite{Li} (at room
temperature) for the first- and second-order
dielectric constants of PbTiO$_3$
(RT). The superscripts $\sigma$ and $\eta$ indicate whether
the measurement is under constant-stress or constant-strain condition.}
\begin{center}\begin{tabular}{lcccc}
 & $\chi^\sigma_{33}$ & $\chi^\eta_{33}$ &
 $\chi_{33}^{(2)\sigma}$ (nm/V) & $\chi_{33}^{(2)\eta}$(nm/V)\\\hline
 model      &67   &  37         & 315 & 82 \\
 experiment  &79   &  33         & -   & -
\end{tabular}\end{center}
\label{tab:dielectric}
\end{table}

In both the free-stress and fixed-strain case, the hysteresis
profile of the static susceptibility shows that $\chi_{33}$
increases with field amplitude for the local minimum at ${\bf\cal
E}<{\bf\cal E}_{\rm c}$ and decreases with increasing field for the
global minimum, which is the only branch in the region above
${\bf\cal E}_{\rm c}$. For each branch, we find a non-linear susceptibility
$\chi^{(2)\sigma}_{33}$ of magnitude $315{\rm nm/V}$ in the stress-free
case.  However, when the strain is clamped, the coercive field
becomes larger than in the stress-free case, and the non-linear
susceptibility is more than two times smaller. In the present
framework, this is not surprising since the change in the
dielectric response is the result of a field-induced change in
structure, and this change is reduced by clamping the strain. In
nonzero field, the susceptibility can be either of these two values
depending on whether the the system is in a single domain
corresponding to the global minimum or to the local minimum, or an
intermediate value if both types of domains are present.

\begin{figure}
\centerline{\epsfig{file=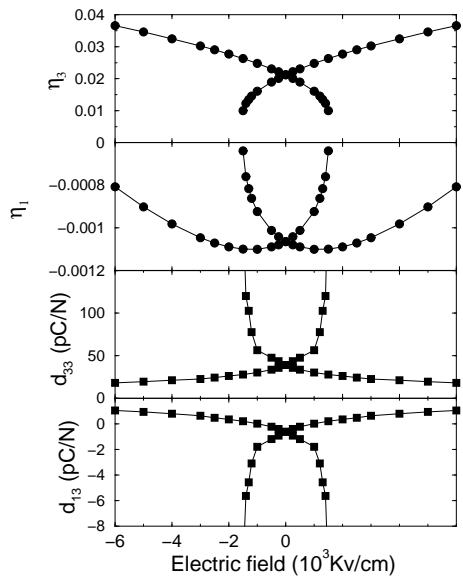,width=6cm}}
\vskip 0.5cm
\caption{The calculated equilibrium strains $\eta_1$ and $\eta_3$ and
the piezoelectric tensor $d_{13}$ and $d_{33}$
as a function of the electric field in PbTiO$_3$.}
\label{fig:PT_piezo}
\end{figure}

Next, we consider the piezoelectric response
(Eq.~\ref{eq:d_field}).  In Fig.~\ref{fig:PT_piezo}, we plot the
equilibrium values of the strains $\eta_1$ and $\eta_3$ as a
function of the electric field along the $z$
 direction. The slopes of these curves give rise to the
 piezoelectric coefficients $d_{13}$ and $d_{33}$ which are plotted
in the lower panels of the same plot. We find $d_{13}=-0.6{\rm
pC/N}$ and $d_{33}= 40{\rm pC/N}$, considerably less than the
room temperature values measured experimentally\cite{Haun} ($-25{\rm pC/N}$ 
and $117{\rm pC/N}$, respectively) and computed from
first principles.\cite{Rabe-Cockayne}
We attribute this primarily to the choice of pseudopotentials, which give a
low value for the ground state tetragonal ratio $c/a$ and in
particular, a value of $a$ almost unchanged from the cubic $a_0$.
However, our calculation does serve to demonstrate the
applicability of our method to the calculation of these
quantities.  In particular, there is to our knowledge no previous
calculation of the nonlinear piezoelectric response.

%--------------------------------------------------------------------
\subsection{Field-Induced Structural Phase Transitions}
\label{sec:SPT}
%-------------------------------------------------------------------
In a single crystal, the relative stability of distorted-structure
phases with polarizations in different directions is expected to
change as an electric field is applied.  In particular, a phase
transition might be induced by applying a sufficiently large field
in a different direction from the polarization of the ground state.
This change in phase, the result of an electric-field induced
rotation of the polarization, may be accompanied by a large change
in strain, manifested as a large piezoelectric response.

This ``polarization rotation" mechanism was proposed in Ref.~\onlinecite{FC} 
to explain the experimentally observed colossal
piezoelectric response to electric fields along [001] of single-crystal 
rhombohedral perovskite alloys such as 
[Pb(Zn$_{1/3}$Nb$_{2/3}$)O$_3$]$_{(1-x)}$--[PbTiO$_3$]$_{x}$
(PZN-PT) with compositions near the rhombohedral--tetragonal morphotropic
phase boundary (R-T MPB),
and has been the subject of continuing
experimental\cite{Noheda01} and theoretical\cite{BGV01} investigation.
Particular attention has focused on the nature of the path
followed by the polarization vector with increasing field strength.
An effective Hamiltonian study of PbZr$_x$Ti$_{1-x}$O$_3$ near the R-T MPB 
\cite{BGV01} showed that with increasing electric field
along [111], the polarization vector of tetragonal PZT rotates
continuously from the tetragonal [001] direction to the rhombohedral
[111] direction through a monoclinic ``$M_A$'' phase\cite{VC01}
with ${\bf P}$ along [$uu$1].
In contrast, for the case of an [001] electric field applied to
rhombohedral PZT, the polarization vector does not simply follow
the return path, but instead follows a discontinuous path
of a kind first discussed by Noheda.\cite{Noheda01}
It first rotates continuously into the $M_A$ phase for small field
strengths, and then jumps discontinuously to a monoclinic ``$M_C$''
phase\cite{VC01} with $\bf P$ along [$u$01] before reaching the
tetragonal structure. 
The calculations show that a large piezoelectric response is expected
for this latter type of path. 

In this section, we apply the full three-dimensional formalism
described in Section~\ref{sec:forma} to study, in a Pb-based
perovskite system, the rotation of polarization by an applied
electric field in the two cases most relevant to enhanced
piezoelectric response near the R-T MPB: (i) application of an
electric field along [111] to a tetragonal system, and (ii)
application of an electric field along [001] to a rhombohedral
system. For (i), we consider tetragonal PbTiO$_3$. For (ii), we
introduce a simple modification of the structural energetics of
PbTiO$_3$ to stabilize a rhombohedral ground-state structure. This
follows the spirit of a view of PZN-PT and PMN-PT as large-strain
PbTiO$_3$-based systems that have been chemically ``engineered" to
make them marginally stable in the rhombohedral
phase.\cite{Cohen-theory}  We do something very similar, but using
a theoretical manipulation that avoids the unnecessary complexities
of the real alloy systems.
 
\subsubsection{Free-energy functional}
Extending the procedure described in Section~\ref{sec:Dielectric} to the
full three-dimensional case, we first 
evaluate the reduced free-energy function
$F(\eta, {\bf P})$ by minimizing $F({\bf R}, \eta, {\bf P})$ with respect
to ${\bf R}$ for a set of selected tetragonal, rhombohedral and
orthorhombic structures. Strains are defined relative
to the cubic structure with the experimental lattice constant
($a_0$ = 7.5 a.u.).
In the range of $\eta$ and ${\bf P}$ of interest, we used a procedure similar
to Sec.~\ref{sec:Dielectric} to fit $F(\eta, {\bf P})$ in a
Landau-Devonshire form: 
\begin{widetext}
\begin{eqnarray}
F(\eta, {\bf P})&=& E_0
+C_1(\eta_1+\eta_2+\eta_3) 
+\frac{1}{2}C_{11}(\eta_1^2+\eta_2^2+\eta_3^2) 
+C_{12}(\eta_2\eta_3+\eta_3\eta_1+\eta_1\eta_2)
+\frac{1}{2} C_{44}(\eta_4^2+\eta_5^2 +\eta_6^2)\nonumber\\
&+&A_{200}(P_x^2+P_y^2+P_x^2)
+A_{400}(P_x^4+P_y^4+P_x^4) 
+A_{220}(P_y^2P_z^2+P_z^2P_x^2+P_x^2P_y^2)\nonumber\\
&+&A_{600}(P_x^6+P_y^6+P_z^6) 
+A_{420}[P_x^2(P_y^4+P_z^4)+P_y^2(P_z^4+P_x^4)
+P_z^2(P_x^4+P_y^4)]
+A_{222}P_x^2P_y^2P_z^2 \nonumber\\
&+&B_{1xx}(\eta_1P_x^2+\eta_2P_y^2+\eta_3P_z^2)
+B_{1yy}[\eta_1(P_y^2+P_z^2)+\eta_2(P_z^2+P_x^2)+\eta_3(P_x^2+P_y^2)]\nonumber\\ 
&+&B_{4yz}(\eta_4P_yP_z+\eta_5P_zP_x+\eta_6P_xP_y) \, .
\label{eq:3Dmodel}
\end{eqnarray}
\end{widetext}
We list the least-squares fitted coefficients in the column denoted by M1
in Table~\ref{tab:coeff3D}. 
\begin{table}[!b]
\caption{Least-squares fitted values of the parameters in
Eq.~(\ref{eq:3Dmodel}). M1, all parameters freely varied;
M2, with the constraint
$A_{222}=0.062{\rm Ha~C}^{-6}{\rm m}^{-12}$ (boldface).
Units are the appropriate combinations of Ha and $(C/m^2)^2$.}
\begin{center}
%\begin{tabular} {c|d|d||c|d|d}
\begin{tabular} {c|c|c||c|c|c}
& M1 & M2 & & M1 & M2\\
\hline
$E_0$ &$-$165.947          &$-$165.947  &$C_1$ &$\phantom{-}$0.168  &$\phantom{-}$0.168             \\
$A_{200}$& $-$0.01 & $-$0.009 &$C_{11}$ &$\phantom{-}$3.829 & $\phantom{-}$3.973 \\
$A_{400}$& $\phantom{-}$0.008 &$\phantom{-}$0.005&$C_{12}$ &$\phantom{-}$1.462  & $\phantom{-}$1.484\\
$A_{220}$& $\phantom{-}$0.015  &$-$0.0007&$C_{44}$ &$\phantom{-}$1.174  & $\phantom{-}$1.218 \\
$A_{600}$& $\phantom{-}$0.003 &$\phantom{-}$0.004&$B_{1xx}$& $-$0.235   & $-$0.234\\
$A_{420}$& $\phantom{-}$0.010 & $\phantom{-}$0.019&$B_{1yy}$& $-$0.048   & $-$0.0525 \\
$A_{222}$& $\phantom{-}$0.009 & {\bf 0}.{\bf 062}&$B_{4yz}$& $-$0.069   & $-$0.068 \\
\end{tabular}\end{center}
\label{tab:coeff3D}
\end{table}

\begin{figure}[!b]
\centerline{\epsfig{file=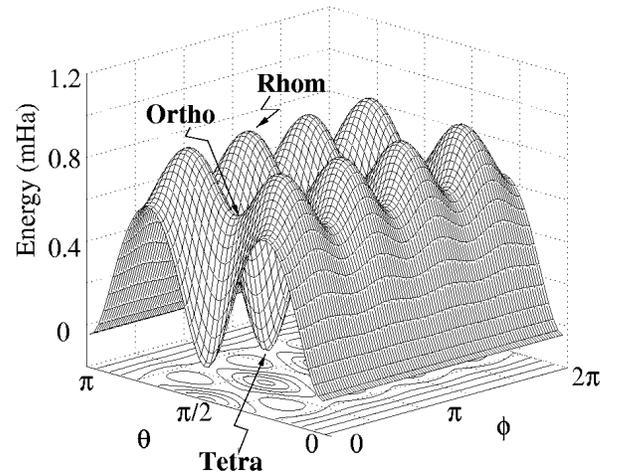,width=8cm}}
\vskip 0.5cm
\caption{Contour plot of $E(\theta, \phi, {\cal E}=0)$ for PbTiO$_3$.
Spherical angles $\theta$ (polar) and $\phi$ (azimuthal) indicate the
direction of $\bf P$.}
\label{fig:PT_phase}
\end{figure}

Using Eq.~(\ref{eq:3Dmodel}), we now consider the energetics of
states with different orientations of the polarization in zero
field. Specifically, we consider $E(\theta,\phi,{\cal E}=0)$,
obtained by fixing the direction of ${\bf P}$ along the direction
specified by spherical angles $\theta$ and $\phi$, relative to the
polar axis $\hat z$, and minimizing $F(\eta, {\bf P})$ with respect
to the strain and the magnitude of $\bf P$.  As shown in
Fig.~\ref{fig:PT_phase}, the tetragonal phase (T) with polarization
along $[001]$ is the global minimum, with a saddle point at the
orthorhombic phase (O) with ${\bf P}\parallel[110]$ and a maximum
at the rhombohedral phase (R), with ${\bf P}\parallel[111]$.  As
shown in Table~\ref{tab:struc_3D}, the structural parameters and the
spontaneous polarizations agree well with the LDA results,
especially for the O and R phases.

\begin{table}
\caption{Comparison between the structural properties of the
tetragonal (T), orthorhombic (O), and rhombohedral (R) phases
of PbTiO$_3$. LDA denotes direct LDA
structural relaxations; M1 and M2 are as in
Table \protect\ref{tab:coeff3D}.
Units of polarization $P$, rhombohedral angle $\alpha$, cell volume
$V$, and phase energies $E$ are C/m$^2$, degrees, Bohr$^3$, and mHa,
respectively.}
%\begin{center}\begin{tabular}{c|ddd}
\begin{center}\begin{tabular}{c|ccc}
              & LDA   & M1    &M2\\\hline
$V_{\rm T}$   &399.9  &402.3  &401.9\\
$c/a$         &1.024    & 1.04  & 1.03\\
$P_{\rm T}$   &0.65  &0.75 & 0.71\\\hline
$V_R$          &398.4&397.3 &398.7\\
$\alpha_{\rm R}$&89.7 & 89.6 & 89.8\\
$P_{\rm R}$ &0.33 &0.32  & 0.34 \\\hline
$V_{\rm O}$ &398.8 &398.1 &399.7\\
$\alpha_{\rm O}$  &89.5  &89.4    & 89.6 \\
$P_{\rm O}$ &0.41  &0.42& 0.44\\\hline
$E_{\rm R}-E_{\rm O}$ & 0.060 & 0.097 &0.064 \\
$E_{\rm O}-E_{\rm T}$ &0.159  & 0.639 & 0.154 \\
\end{tabular}\end{center}
\label{tab:struc_3D}
\end{table}

From this table, it can also be seen that the energy differences
between the T, O and R phases are quite small. For this reason, the
parameters obtained by a global least-squares minimization do not
accurately reproduce the LDA values. In particular, the energy
difference between the T and R phases is seen to be much larger
than the LDA result. However, these features of the energy surface
are crucial to the physics of the structural phase transitions.
Therefore, we adjusted the fitting procedure to reproduce these
relative energies accurately while using the least-squares
procedure for the best overall fit to the remaining data, as
follows. Rather than introduce additional parameters by including
higher-order terms, we ``tune" one parameter while determining the
other 13 parameters by standard least-squares minimization, and
choose the value for the single tuned parameter that yields the
most accurate values for {\it both} the O-T and O-R energy
differences.
$A_{222}$ proves to be the best choice for the tuning parameter,
and with $A_{222}$= 0.062 and the other parameters as given in the
column denoted by M2 of Table \ref{tab:coeff3D}, both the O-T and
O-R energy differences as well as the structural parameters and
spontaneous polarizations of all three phases are in excellent
agreement with the LDA results, as shown in the last column of
Table~\ref{tab:struc_3D}. Therefore, this set of parameters was used
in the following calculations.

\subsubsection{Engineering a Rhombohedral Structure for PbTiO$_3$}

In previous first-principles investigations of PbTiO$_3$, it has
been observed that the strain coupling is responsible for stabilizing
the tetragonal ground state structure.\cite{Cohen90} In the simple
cubic lattice, the lowest-energy structure has polarization along
[111], corresponding to a rhombohedral symmetry, while the energy
of the optimal state with polarization along [001] is higher.
However, when the lattice is allowed to relax, the energy gain from
strain coupling in the tetragonal structure is much larger than the
gain in the rhombohedral structure, leading to the observed
reversal of stability.  In each case, the energy gain from strain
coupling increases as the relevant elastic constant decreases. So,
if it were possible to decrease the shear modulus $C_{44}$, there
would be a critical value below which the rhombohedral state would
be most stable.

\begin{figure}
\centerline{\epsfig{file=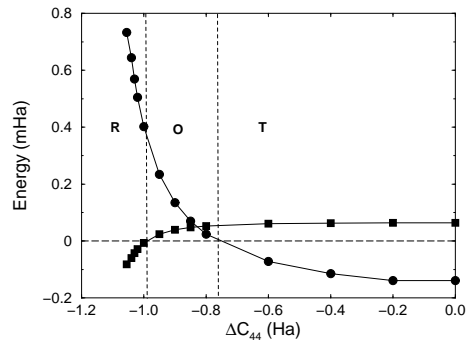,angle=270,width=6cm}}
\caption{The energies of the rhombohedral (square) and tetragonal
(circle) phases relative to the orthorhombic phase (chosen as the
zero of energy) as a function of the tunable shear modulus
$\Delta C_{44}$ of PbTiO$_3$, calculated using Eq.~(\ref{eq:Tuneshear}).
The ranges of $\Delta C_{44}$ in which the tetragonal,
orthorhombic and rhombohedral phases are most stable are separated
by vertical dashed lines and indicated by T, O and R respectively.}
\label{fig:phase_Model}
\end{figure}

Within Eq.~(\ref{eq:3Dmodel}), the modification of $C_{44}$ can be
implemented by the inclusion of a tunable shear elastic term
\begin{equation}
\widetilde{F}({\eta},{\bf P}) = F({\eta},{\bf P})+\frac{1}{2}\Delta C_{44}
(\eta_4^2+\eta_5^2 +\eta_6^2)
\label{eq:Tuneshear}
\end{equation}
where $\Delta C_{44}$ = 0 corresponds to PbTiO$_3$ with its natural
shear elastic modulus. Using Eq.~(\ref{eq:Tuneshear}), we compute the
zero-field energy for the optimal tetragonal, rhombohedral and
orthorhombic phases as a function of $\Delta C_{44}$. This yields
the phase sequence shown in Fig.~\ref{fig:phase_Model} with the T
and R phases separated by a sliver of an orthorhombic phase. This
phase sequence is very reminiscent of that of the
Pb(Zn$_{1/3}$Nb$_{2/3}$)O$_3$--PbTiO$_3$ system\cite{Cox} with the
tunable parameter being the proportion of PbTiO$_3$.  The stability
of the orthorhombic phase reflects the importance of the
sixth-order terms in Eq.~(\ref{eq:3Dmodel}), as in a fourth-order
model only tetragonal and rhombohedral structures are possible
minima.

\begin{figure}
\centerline{\epsfig{file=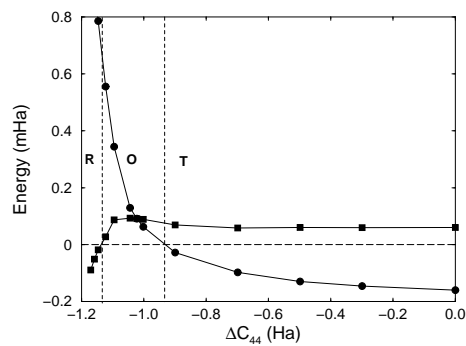,angle=270,width=6cm}}
\caption{Same as Fig.~\ref{fig:phase_Model} but calculated
using a direct LDA approach. As in
Fig.~\ref{fig:phase_Model}, an orthorhombic window appears, though
the phase boundaries are slightly shifted.}
\label{fig:phase_LDA}
\end{figure}

To check that the observed phase sequence is not an artifact of
our fit, we have computed the structural parameters and energies of 
the tetragonal, orthorhombic and rhombohedral phases as a function
of $\Delta C_{44}$ through direct LDA calculations. For consistency with 
Eq.~(\ref{eq:Tuneshear}), we implement the adjustment of the shear
modulus as an additional applied stress
\begin{equation}
\Delta\sigma_i = - \Delta C_{44}\eta_i  
\end{equation}
where $\sigma_i$ (with $i=4,5,6$) are the shear stress components in Voigt
notation. The results, given in Fig.~\ref{fig:phase_LDA}, show the same
T-O-R phase sequence as Fig.~\ref{fig:phase_Model}. 
While the T-O and O-R phase boundaries are slightly
shifted, the width of the orthorhombic window is comparable to that in
Fig.~\ref{fig:phase_Model}. Thus, in the following, using
Eq.~(\ref{eq:Tuneshear}) with
a particular value of $\Delta C_{44}$, we expect results which would 
reflect a direct LDA calculation, though perhaps with a slightly different 
$\Delta C_{44}$.

\subsubsection{Electric-field-induced phase transitions}

In a single crystal, the relative stability of phases with polarizations
in different directions is expected to change as an electric field is
applied. In particular, a phase transition might be induced by applying
a sufficiently large field in a different direction from the polarization 
of the ground state. Here, we consider two such cases: tetragonal PbTiO$_3$ 
in an electric field along [111], and rhombohedral ``PbTiO$_3$," stabilized
by a nonzero value of $\Delta C_{44}$, in an electric field along [001].

\begin{figure}
\centerline{\epsfig{file=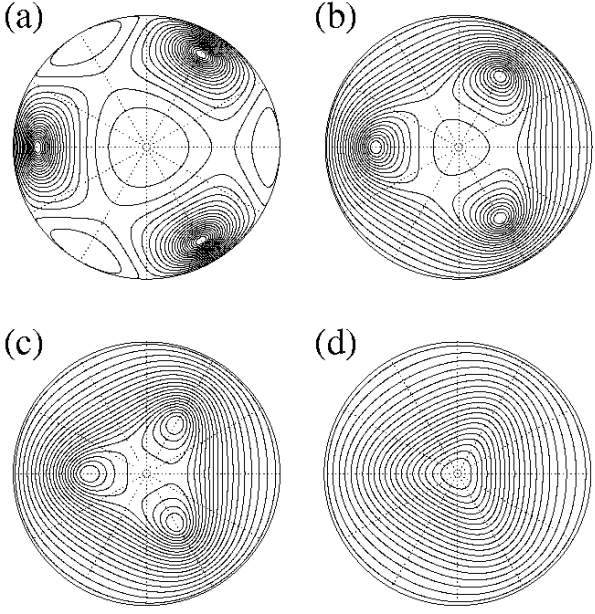,width=8cm}}
\caption{Contour maps of $E(\theta,\phi,{\cal E})$ on
the upper hemisphere $0\leq\theta\leq\frac{1}{2}\pi$ for an
electric field of magnitude ${\cal E}$ applied along the
pseudocubic [111] direction to tetragonal PbTiO$_3$. The contour is
equally spaced in $log(E-E_{\rm min}+\delta)$, where $E_{\rm min}$
is the global minimum and $\delta$ is a small offset.  The central
axis points along the [111] direction. (a)-(d) correspond to
electric fields of 0, 0.86, 1.73, and 3.46$\times 10^3$kV/cm,
respectively.}
\label{fig:E111}
\end{figure}

First, we consider tetragonal PbTiO$_3$ ($\Delta C_{44}$ = 0) in an
electric field along [111], which tends to favor a rhombohedral
direction for the polarization.
To investigate the evolution of various phases with ${\cal E}_{111}$,
where ${\cal E}_{111}$ is the magnitude of the electric field, we
perform the minimization in two steps. First, we transform the Euclidean
coordinates ($P_{x}$,$P_{y}$,$P_{z}$) into spherical coordinates
($P$,$\theta$,$\phi$) and compute 
\begin{equation}
E(\theta,\phi,{\cal E}_{111})= \min_{
P,\eta}[F(\eta,{\bf P})- 
{\cal E}_{111}\,(P_{x}+P_{y}+P_{z})/\sqrt{3}]
\label{eq:E111} 
\end{equation}
Then, we locate the minima on the sphere of polarization directions 
parametrized by $\theta$ and $\phi$.

The evolution of the phase stability can be readily displayed by the
contour plots of $E(\theta,\phi,{\cal E}_{111})$ shown in
Fig.~\ref{fig:E111}.  At zero electric field,
the tetragonal structure appears as a three-fold degenerate energy 
minimum in the hemisphere shown. As ${\bf {\cal E}}_{111}$ increases,
the minima migrate from the tetragonal positions along the lines corresponding
to the monoclinic $M_A$ phase (three-fold degenerate) and eventually reach
the rhombohedral point at the center of the hemisphere. 

\begin{figure}
\centerline{\epsfig{file=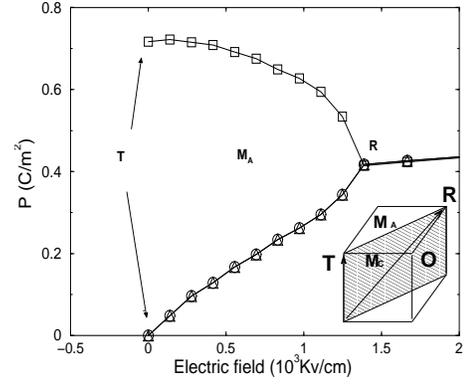,width=6cm}}
\caption{The Cartesian component  $P_x$ (circle), $P_y$ (triangle),
$P_z$ (square) of the polarization as a
function of the magnitude of the electric field applied along the [111]
pseudocubic direction in PbTiO$_3$. Inset shows the polarization path.}
\label{fig:Pol_111}
\end{figure}

Figure \ref{fig:Pol_111} shows how the polarization components
of tetragonal PbTiO$_3$ evolve with the amplitude of ${\bf {\cal
E}}_{111}$. At ${\bf {\cal E}}_{111}=0$, the only non-zero
component is $P_z$. As ${\bf {\cal E}}$ increases, $P_x=P_y$ grow
while $P_z$ slowly decreases. The structure thus enters the $M_A$
monoclinic phase.  When ${\bf {\cal E}}$ reaches $1.4\times
10^3{\rm kV/cm}$, the three components merge and the system enters
the rhombohedral phase where the polarization vector points along
the pseudocubic [111] direction. While rotating, the polarization
vector remains in the (110) plane, as shown in the inset of
Fig.~\ref{fig:Pol_111}.

\begin{figure}
\centerline{\epsfig{file=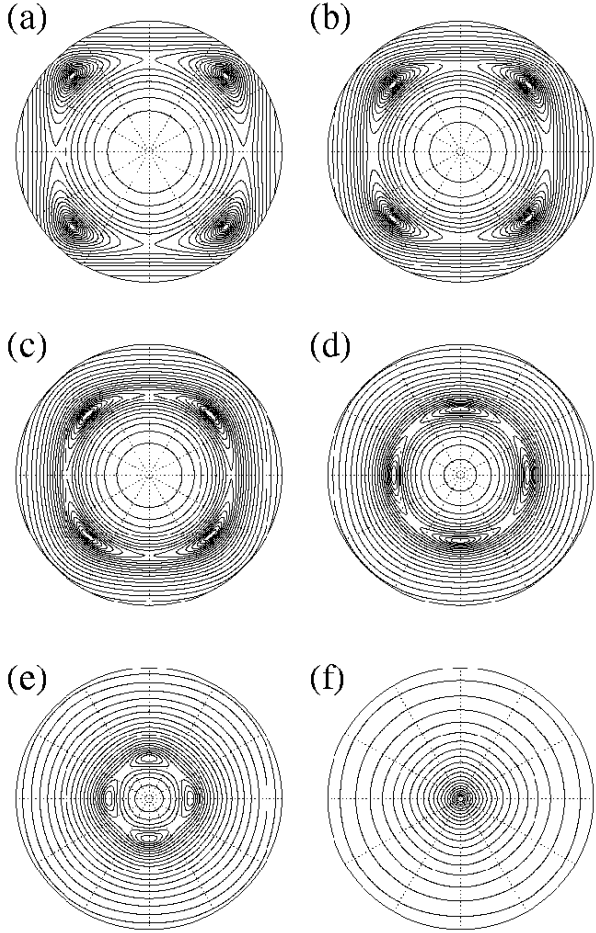,width=8cm}}
\caption{Contour map of the free energy
(upper hemisphere) when an electric field is applied to
rhombohedral ``PbTiO$_3$'' ($\Delta C_{44} = -1.1{\rm Ha}$)
along the pseudocubic [001] direction.
The central axis corresponds to one of the tetragonal directions.
(a)--(f) correspond to electric field magnitudes of 0, 1.4,
2.8, 7, 14, and $19\times10^3{\rm kV/cm}$, respectively.}
\label{fig:E001}
\end{figure}
 
Next, we consider rhombohedral ``PbTiO$_3$" with $\Delta C_{44}=-1.1{\rm Ha}$
(see Fig.~\ref{fig:phase_Model})
in an electric field along the [001]
direction, which tends to favor a tetragonal direction for the polarization. 
The analogue of Eq.(~\ref{eq:E111}) is
\begin{equation}
E(\theta,\phi,{\cal E}_{001})= \min_{P,\eta}[\widetilde{F}(
{\eta},{\bf P})- {\cal E}_{001}\,P_{z}]
\end{equation}
The energy contour plot in this case is shown in
Fig.~\ref{fig:E001}. In zero electric field, the system is in a
rhombohedral phase with an eight-fold degenerate minimum. For small
nonzero ${\cal E}_{001}$, the energy minima correspond to a $M_A$
phase as shown in Fig.~\ref{fig:E001}(b-c) where there are four
degenerate minima lying in the (110) plane.  At a critical value of
${\cal E}_{001}$, the energy minima jump to four-fold points in the
(100) plane, as can be seen in Fig.~\ref{fig:E001}(d). The four
minima then move smoothly towards the [001] axis, finally merging
to yield the tetragonal phase.

\begin{figure}
\centerline{\epsfig{file=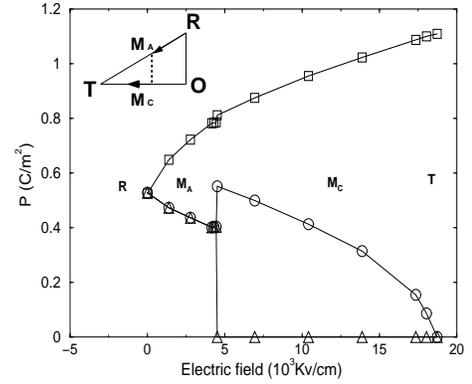,width=6cm}}
\caption{Same as Fig.~\ref{fig:Pol_111}, for an electric field applied
along the [001] pseudocubic direction in rhombohedral ``PbTiO$_3$" obtained
with $\Delta C_{44} = -1.1{\rm Ha}$.}
\label{fig:Pol_001}
\end{figure}

Figure \ref{fig:Pol_001} shows how the polarization components of
rhombohedral ``PbTiO$_3$" evolve with the amplitude of ${\bf {\cal
E}}_{001}$.
Under zero applied electric field, the polarization vector starts
along the pseudocubic [111] direction ($P_x=P_y=P_z=0.56$C/m$^2$). 
As ${\bf {\cal E}}_{001}$ increases, the structure enters an $M_A$ phase
in which $P_x$ and $P_y$ remain equal, but become less than $P_z$. 
$P_x$ and $P_y$ keep dropping
until $P_y$ shows a sudden jump to zero at 
around $4.5\times10^3{\rm kV}/{\rm cm}$. At the same time,  
both $P_x$ and $P_z$ exhibit an upward jump 
in their values. The new phase corresponds to a different monoclinic
phase denoted by $M_C$.
The structure remains in the $M_C$ phase until $P_x$ also drops to zero 
at around $19\times10^3{\rm kV}/{\rm cm}$, yielding 
a tetragonal phase. As the field increases further, $P_z$ 
continues to increase smoothly.

In this section, we have seen that a small modification of the
structural energetics of PbTiO$_3$ can yield a complex polarization 
path quite similar to that proposed by Noheda\cite{Noheda01} and 
observed in simulations of PZT.\cite{BGV01}
Additional calculations, for example of the lattice parameters as a function
of electric field, may assist in achieving a direct experimental observation
of this behavior. In addition, further exploration within this framework may
suggest ways to produce and control complicated polarization paths in real
systems.

\section{Summary}
\label{sec:Summary}
In this paper, we have introduced
a formalism for computing the structural response of an insulating
system to a static homogeneous macroscopic electric field. We have shown that,
in the presence of an electric field, 
the thermodynamic potential $E({\bf R},\eta,{\cal E})$  can be minimized
by introducing a related thermodynamic potential $F({\bf R},\eta,{\bf P})$ in which
the polarization ${\bf P}$ is treated as a fundamental variable. 
Corresponding to each polarization ${\bf P}$,
the equilibrium values for the internal coordinates ${\bf R}$ and $\eta$
as well as the minimum of this energy functional can be
computed. Consequently, one arrives at an energy functional that only
depends on ${\bf P}$ and where
the effect of a homogeneous electric field can be 
treated exactly by adding a linear term $-{\cal E}\cdot{\bf P}$ to this functional.

In practice, when $E({\bf R},\eta,{\cal E})$ 
is expanded to first order in ${\cal E}$,  
the minimization is reduced to one 
over the internal coordinates constrained by a fixed 
polarization computed at zero electric field.  
We have implemented a  minimization scheme in the framework of a 
modified DFPT, using a consistent discretization formula that was developed
for the response to an electric field. Consequently, the computed 
response is compatible with the Berry-phase polarization, which is a 
central quantity in the formalism. 

It is important to note that the present $i=1$ theory is most useful for
systems in which the response to an electric field is dominated
by the changes in atomic coordinates and strains rather than by
electronic polarization.
Ferroelectric and nearly-ferroelectric materials are among the
best examples.  We therefore look forward to future 
applications of our new approach for a variety of purposes, 
for example, ferroelectric alloys and ferroelectric
superlattices. Applying the method to the so called 
``high-K materials'' to study their dielectric properties in the
presence of an applied electric field also appears to be 
a promising direction.  

Though the higher-order (say $i=2$) theory requires higher ($\ge3$) 
order energy derivatives, this does not preclude its application.
As mentioned in
Appendix~\ref{app:higherorder}, it is possible to approximate certain
response quantities that are related to the third derivatives by constant
values from a single structure, if they show only small variations 
within the range of the polarization studies. Systems that
may satisfy such a condition will be the subject of further investigation.  

\acknowledgments
This work was supported by ONR Grants N0014-97-1-0048 and
N00014-00-1-0261. The work of K.M.R. was performed in part at the Aspen
Center for Physics. We would like to thank X. Gonze for his interest in the work and
valuable discussions on the {\tt Abinit} code. We acknowledge
Ph. Ghosez and M. Veithen for their help on the FHI pseudopotentials. 

\appendix
\section{Second-Order Expansion Formalism}
\label{app:higherorder}

This appendix presents the formalism in Section~\ref{sec:forma} 
for truncation of the sum in Eq.~(\ref{eq:Eexp}) at $i=2$, that is,
at second order in the electric field ${\cal E}$. At this order, the
thermodynamic potential $E({\bf R}, \eta, {\cal E})$
is replaced by $E_2({\bf R}, \eta, {\cal E})$, which is the sum of the 
first three terms in Eq.~(\ref{eq:Eexp}). 

We recall the definition of the dielectric susceptibility tensor 
\begin{equation}
\chi_{\alpha\beta}({\bf R},\eta,{\bf {\cal E}}) = 
-\frac{1}{\epsilon_0}\frac{\partial^2{E({\bf R},\eta,{\bf {\cal E}})}}{\partial{\bf {\cal
E}}_\alpha\partial{\bf {\cal E}}_\beta} =
\frac{1}{\epsilon_0}\frac{\partial{ P_\alpha({\bf R},\eta,{\bf {\cal E}})}}{\partial{\bf
{\cal E}}_\beta} \, .
\label{eq:Chidef}
\end{equation}
Therefore, we can write 
\begin{widetext}
\begin{equation}
E_2({\bf R},\eta,{\cal E})=E({\bf R},\eta,{\cal
E})-\sum_{\alpha}P_\alpha({\bf R},\eta,0){\cal E}_\alpha
-\frac{\epsilon_0}{2}\sum_{\alpha\beta}{\bf {\cal E}}_\alpha {\bf {\cal
E}}_\beta \chi_{\alpha\beta}({\bf R},\eta,0)
\label{eq:E2}
\end{equation} 
and
\begin{equation}
P_{2,\alpha}({\bf R},\eta,{\cal E})=P_\alpha({\bf R},\eta,0) +\epsilon_0\sum_{\beta} {\bf {\cal
E}}_\beta \chi_{\alpha\beta}({\bf R},\eta,0) \, .
\label{eq:P2}
\end{equation}

The computation of $F({\bf P})$ (Eq.~\ref{eq:Fdef}) 
for a given ${\bf P}$ proceeds by the  
minimization of $E({\bf
R},\eta,{\bf {\lambda}})+ \lambda\cdot{\bf P}$ 
following the procedure in Section IIIA. This 
involves computing the derivatives
\begin{equation}
\frac{\partial E_2({\bf R},\eta,{\lambda})}{\partial R_{i\gamma}} = 
\frac{\partial {E({\bf R},\eta,0)}}{\partial R_{i\gamma}}-
\sum_{\alpha}{\frac{\partial P_\alpha({\bf R},\eta,0)}{\partial R_{i\gamma}}\lambda_\alpha} -
\frac{\epsilon_0}{2}\sum_{\alpha\beta}\lambda_\alpha \lambda_\beta
{\frac{\partial {\chi}_{\alpha\beta}({\bf R},\eta,0)}{\partial
R_{i\gamma}}} \, ,
\label{eq:grad_R}
\end{equation}
\begin{equation}
\frac{\partial E_2({\bf R},\eta,{\lambda})}{\partial\eta_\mu}=\frac{\partial {E({\bf R},\eta,0)}}{\partial{\eta}_\mu}-\sum_{\alpha}{\frac{\partial P_\alpha({ \bf R},\eta,0)}{\partial {\eta_\mu}}} \lambda_\alpha -
\frac{\epsilon_0}{2}\sum_{\alpha\beta} \lambda_\alpha \lambda_\beta
\frac{\partial {\chi}_{\alpha\beta}({\bf R},\eta,0)}{\partial {\eta}_\mu}\, ,
\label{eq:grad_eta}
\end{equation}
\begin{equation}
\frac{\partial E_2({\bf R},\eta,{\lambda})}{\partial\lambda_\alpha} =
-P_\alpha({\bf R},\eta,0) -
\epsilon_0\sum_{\beta}\chi_{\alpha\beta}({\bf R},\eta,0)\lambda_\beta +
P_\alpha \, .
\label{eq:grad_lambda}
\end{equation}
\end{widetext}
These are related to the corresponding derivatives in the $i=1$ case
(Eq.~\ref{eq:mincond}) by the addition of terms one order higher 
in $\lambda$.
From Eq.~(\ref{eq:grad_lambda}), we see that at this order 
${\bf P}({\bf R},\eta,\lambda)$ includes an electronic contribution
$\epsilon_0\sum_{\beta}\chi_{\alpha\beta}({\bf R},\eta,0)\lambda_\beta$.
The effective forces and stresses (Eqs.~\ref{eq:grad_R} and \ref{eq:grad_eta})
involve the derivatives of $\chi$ with respect to 
$\bf R$ and $\eta$.
While these are in principle obtainable from the 2n+1 theorem, they
are not routinely calculated in current DFPT codes.
For cases where the lattice contribution to $\bf P$ dominates, it
is reasonable however to approximate the 
$\bf R$ and $\eta$ dependence of $\chi$ by evaluating it at the
zero-field equilibrium structure. 
A more accurate but still practical approximation would include
the first order changes with respect to $\delta {\bf R}$ and $\delta \eta$,
with the derivatives computed through a finite difference approach.

\section{Multiband discretization formula}
\label{app:multi}
In Sec.~\ref{sec:Method-compatibility} we presented a finite-difference
formula, Eq.~(\ref{eq:vk-finite}), representing the derivative
$i\partial \vert u_k\rangle/\partial k$ in the single-band
1D case.  In this Appendix we generalize the derivation in order to
obtain a corresponding formula for the multiband 3D case.

The general expression for the electronic polarization in 3D
is easily reduced to a sum of 1D Berry phases over strings of $\bf k$
points.\cite{KVBerry}  We can write
\begin{equation}
{\bf P}=\frac{1}{VN_k}\sum_{k_\perp}\sum_\alpha {\bf R}_\alpha P_\alpha(k_\perp)
\label{eq:P-multi}
\end{equation} 
where $V$ is the cell volume, $\alpha$ labels the three primitive
real-space lattice vectors ${\bf R}_\alpha$ conjugate to the
primitive reciprocal-space vectors ${\bf G}_\alpha$, and $k_\perp$ runs
over a 2D mesh of $N_k$ positions in the reciprocal-space
directions perpendicular to $\alpha$.  The contribution from
the string ${\cal S}(k_\perp)$ of ${\bf k}$-points running
parallel to ${\bf G}_\alpha$
at a given $k_\perp$ is
\begin{equation}
P_\alpha(k_\perp)=-\frac{fe}{2\pi}
\sum_{{\bf k}\in S(k_\perp)} {\rm Im} \ln \det M^{(\bf k,k+b)} \, ,
\label{eq:P-reduce}
\end{equation}
where $f=2$ for spin,
\begin{equation}
M^{(\bf k,k+b)}_{mn}= {\langle u_{m{\bf k}}\vert u_{n,{\bf k+b}}\rangle}
\label{eq:overlapM}
\end{equation} 
is the overlap matrix formed of inner products
between Bloch orbitals on neighboring ${\bf k}$-points on the string,
${\bf b}$ is the separation between neighboring points on
the string, and $m$ and $n$ run over the
occupied valence bands.  Equation (\ref{eq:P-reduce}) is
essentially the multi-band generalization of Eq.~(\ref{eq:berry-1d})
of Sec.~\ref{sec:Method-compatibility}.

For the remainder of this Appendix, we drop the 3D notation and
start from the 1D version
\begin{equation}
P=-\frac{fe}{2\pi} \sum_k {\rm Im} \ln \det M^{(k,k+b)}
\label{eq:P-1d}
\end{equation}
of Eq.~(\ref{eq:P-reduce}), and correspondingly for
Eq.~(\ref{eq:overlapM}).  Our task is to compute the
variation $\delta P$ arising from the first-order variations
of the wavefunctions in Eq.~(\ref{eq:P-1d}).  Focusing
on a single wavevector $k$ and its neighbor $k'=k+b$
and letting $M=M^{(k,k')}$, our central task is clearly to
compute the first-order variation of the phase
\begin{equation}
\phi = {\rm Im}\ln \det M \, .
\end{equation} 
Using
\begin{equation}
\det M = \sum_{\hat{\rm p}}(-1)^{\hat{\rm p}} \prod_{n}{\langle u_{nk}\vert
u_{{\hat{\rm p}}(n)k'}\rangle} \, ,
\end{equation}  
where $\hat{\rm p}$ runs over all possible permutations
among the occupied bands, the change in this phase from a
first-order change in the wavefunctions at $k$ is
\begin{equation}
\delta \phi = {\rm Im} \frac{\delta \det M}{\det M_0}
\label{eq:delta-phi}
\end{equation}
where
\begin{eqnarray}
\delta \det M &=& \sum_{\hat{\rm p}}(-1)^{\hat{\rm p}}\sum_n \langle \delta
u_{nk}\vert u_{\hat{\rm p}(n)k'}\rangle
\prod_{m\neq n} \langle u_{mk}\vert u_{\hat{\rm p}(m)k'}\rangle
\nonumber \\
&=& \sum_{\hat{\rm p}}(-1)^{\hat{\rm p}}\sum_n \langle \delta
u_{nk}\vert u_{\hat{\rm p}(n)k'}\rangle
\prod_{m\neq n} M_{0,m\hat{\rm p}(m)} \, .
\label{eq:mess}
\end{eqnarray}
Here $M_0$ is the matrix $M$ evaluated before variation
of the wavefunctions.

Unfortunately, Eq.~(\ref{eq:mess}) does not lend itself to simple
evaluation.  However, we can reduce Eq.~(\ref{eq:mess}) to a trivial
form as follows.  Consider a linear transformation
\begin{equation}
\vert\tilde u_{nk'}\rangle \equiv \sum_{m}A_{mn}\vert u_{mk'}\rangle
\label{eq:trans}
\end{equation}
among the occupied states at $k'$, where $A$ is a non-singular (but
not necessarily unitary) matrix.  Letting
$\widetilde{M}_{mn}=\langle u_{mk} \vert \tilde u_{nk'}\rangle$,
it follows that $\widetilde{M}=MA$ and thus
$\det(\widetilde{M})=\det(M)\det(A)$.  Since $A$ is a constant matrix,
\begin{equation}
\delta\ln\det\widetilde{M}=\delta\ln\det M \, .
\end{equation}
We thus have the freedom
to evaluate Eqs.~(\ref{eq:delta-phi}) and (\ref{eq:mess})
with the substitutions $M\rightarrow\widetilde{M}$,
$M\rightarrow\widetilde{M}_0$ and $u_{mk'}\rightarrow
\tilde u_{mk'}$, where $\widetilde{M}=MA$ and
$\widetilde{M}_0=M_0A$, for arbitrary $A$.

The obvious choice is $A=M_0^{-1}$.  We then find that the only permutation
that survives in Eq.~(\ref{eq:mess}) is the identity and the
denominator of Eq.~(\ref{eq:delta-phi}) becomes unity, so that
\begin{equation}
\delta\phi = {\rm Im} \sum_n \langle \delta u_{nk}\vert
\tilde{u}_{nk'}\rangle
\label{eq:delta_phi}
\end{equation}
where
\begin{equation}
\vert\tilde{u}_{nk'}\rangle = \sum_{m}(M_0^{-1})_{mn}\vert
u_{mk'}\rangle \, .
\label{eq:unk'}
\end{equation}
Eq.~(\ref{eq:delta_phi}) can also be written neatly as
\begin{eqnarray}
\delta\phi = {\rm Im}\, {\rm Tr}(\delta M\cdot M_0^{-1}) \, .
\end{eqnarray}
Carrying out similar manipulations for the connection between
$k$ and $k-b$, we can define
\begin{equation}
\vert v_{nk}\rangle \equiv
\frac{i}{2b}\left( \vert
\tilde{u}_{n,k+b}\rangle  - \vert \tilde{u}_{n,k-b}\rangle  \right)
\label{eq:vnkmul}
\end{equation}
which becomes the finite-difference representation of
$i\partial \vert u_{nk}\rangle/\partial k$ in the multiband case,
analogous to Eq.~(\ref{eq:vk-finite}).  It is easy to
check the orthogonality of the $v_{nk}$ to the occupied subspace,
\begin{equation}
\langle u_{nk}\vert v_{mk}\rangle = \frac{i}{2b}
(\delta_{nm} - \delta_{nm}) = 0 \, ,
\label{eq:uvortho}
\end{equation}
thus removing the need for explicit application of a conduction-band
projector onto the $\vert v_{nk}\rangle$ when computing the
right-hand side of Eq.~(\ref{eq:field-Stern}).
Since $\langle u_{mk} \vert \tilde u_{nk'} \rangle=\delta_{mn}$, we
can think of $\vert\tilde{u}_{nk'}\rangle$ defined in
Eq.~(\ref{eq:unk'}) as a phase-aligned and amplitude-corrected
``partner''  to $\vert u_{nk}\rangle$ formed from the occupied subspace
at $k'$, and $\vert v_{nk}\rangle$ is proportional to the
difference between the ``partners'' at $k+b$ and $k-b$.

\newpage

Finally, the variation of Eq.~(\ref{eq:P-1d}) becomes
\begin{equation}
\delta P = \frac{feb}{\pi}\sum_k
{\rm Re}\,\langle \delta u_{nk}\vert v_{nk}\rangle
\label{eq:varP}
\end{equation}
in analogy with Eq.~(\ref{eq:dP_finite}).

Our implementation of this scheme into {\tt ABINIT} is based on
Eqs.~(\ref{eq:delta_phi}-\ref{eq:varP}) above.

\end{document}